\documentclass[twocolumn]{aastex631}

\usepackage{amsmath}
\usepackage{enumitem}

\shorttitle{Hierarchical SN Classification}
\shortauthors{Villar et al.}

\begin{document}

\title{The Impact of Host Galaxy Properties on Supernova Classification with Hierarchical Labels}

\author[0000-0002-5814-4061]{V.~Ashley Villar}
\affil{Center for Astrophysics \textbar{} Harvard \& Smithsonian, 60 Garden Street, Cambridge, MA 02138-1516, USA}
\affil{The NSF AI Institute for Artificial Intelligence and Fundamental Interactions}

\author[0000-0001-6395-6702]{Sebastian~Gomez}
\affil{Space Telescope Science Institute, 3700 San Martin Dr, Baltimore, MD 21218, USA}

\author[0000-0002-9392-9681]{Edo~Berger}
\affil{Center for Astrophysics \textbar{} Harvard \& Smithsonian, 60 Garden Street, Cambridge, MA 02138-1516, USA}
\affil{The NSF AI Institute for Artificial Intelligence and Fundamental Interactions}

\author[0000-0003-4906-8447]{Alex~Gagliano}
\affil{Center for Astrophysics \textbar{} Harvard \& Smithsonian, 60 Garden Street, Cambridge, MA 02138-1516, USA}
\affil{The NSF AI Institute for Artificial Intelligence and Fundamental Interactions}

\begin{abstract}
With the advent of the Vera C. Rubin Observatory, the discovery rate of supernovae (SNe) will surpass the rate of SNe with real time spectroscopic followup by three orders of magnitude. Accurate photometric classifiers are essential to both select interest events for followup in real time and for archival population-level studies. In this work, we investigate the impact of observable host galaxy information on the classification of SNe, both with and without additional light curve and redshift information. We find that host galaxy information alone can successfully isolate relatively pure ($>90$\%) samples of Type Ia  SNe with or without redshift information. With redshift information, we can additionally produce somewhat pure ($>70$\%) samples of Type II SNe and superluminous supernovae. Additionally with redshift information, host galaxy properties do not significantly improve the accuracy of SN classification when paired with complete light curves. In the absence of redshift information, however, galaxy properties significantly increase the accuracy of photometric classification. As a part of this analysis, we present the first formal application of a new objective function, the weighted hierarchical cross-entropy, to the problem of supernova classification. This objective function more naturally accounts for the hierarchical nature of supernova classes and, more broadly, transients. Finally, we present a new set of SN classifications for the Pan-STARRS Medium Deep Survey of SNe that lack spectroscopic redshift, increasing the full photometric sample to $>$4400 events.
\end{abstract}


\section{Introduction} \label{sec:intro}
Early classification of extragalactic transients, in particular supernovae (SNe), is paramount to enabling multi-wavelength and spectroscopic analysis in real time. Currently, $\sim10$ percent of SNe receive a spectroscopic classification, the traditional means for understanding the underlying nature of a SN. The fraction of SNe which remain spectroscopically unclassified will significantly grow with the upcoming Legacy Survey of Space and Time (LSST) conducted by Vera C. Rubin Observatory and expected to commence in 2025. LSST will discover over one million SNe annually \citep{kessler2019models}; without additional spectroscopic resources, $\lesssim0.1\%$ of all SNe will have spectroscopic followup. Large-scale spectroscopic followup campaigns, such as 4MOST \citep{swann20194most}, will increase this to $\sim1$\%. As a result of this mismatch in discovery vs followup rates, the time-domain community has placed significant emphasis on \textit{photometric} classifiers in the years leading up to LSST. 

Photometric classifiers aim to classify SNe into their historically spectroscopic classes based on photometric data alone. The taxonomy of SNe has actively evolved as increasingly large samples unveil new diversity in SNe observables, leading to a branching hierarchical structure. Broadly speaking, Type II SNe are those which show spectroscopic signatures of hydrogen near the peak of their optical emission. Within this class, the spectra of Type IIP/L SNe have broad H (typically with P-Cygni profiles), while Type IIn SNe are dominated by narrow H emission. In contrast, Type I SNe lack said features, with Type Ib SNe showing signs of helium and Type Ic SNe lacking signs of either. Type Ibc and II SNe arise from the core-collapse of massive stars. Type Ia SNe, arising from thermonuclear explosions of white dwarfs, lack H and He in their optical spectra but show strong signatures of Si II near peak. A broad overview of the hierarchical SN taxonomy is presented in \cite{gal2016observational}.

Some photometric classification methods can be conducted in real time (e.g., \citealt{muthukrishna2019rapid, moller2020supernnova,carrasco2021alert, qu2022photometric, gagliano2023first}) while others rely on complete light curves (e.g., \citealt{hosseinzadeh2020photometric,boone2021parsnip}) for feature extraction. It can be especially challenging to identify and measure useful SN features in the early photospheric phases of the light curves. However, prompt identification of SNe is key to capture early observational phenomena (e.g., flash spectroscopy for young core-collapse SNe; \citealt{khazov2016flash, bruch2021large,jacobson2024final}), to guide multiwavelength follow up and to optimize spectroscopic resources. In order to perform this early classification, \textit{all} available information must be utilized, including the \textit{contextual} information provided by the host galaxy of the transient. 

Specifically, host galaxy information is known to correlate with transient properties. For example, while thermonuclear (Type Ia) SNe are broadly observed across all galaxy types, core-collapse supernovae (CCSNe) occur only in galaxies with recent or ongoing star formation \citep{leaman2011nearby,hakobyan2012supernovae,childress2013host}. Type Ib and Ic SNe, specific substypes of stripped-envelope SNe (SESNe), prefer slightly more massive (and typically higher metallicity) host galaxies compared to their hydrogen-rich counterparts of Type II(b) SNe \citep{schulze2021palomar}. Similarly, SESNe are over-represented in disturbed galaxies compared to their H-rich counterparts \citep{habergham2012central}. Rare CCSN classes seem to show stronger and more exotic preferences. Both Type Ic-BL (high-energy, ``broad-lined" events) and H-poor superluminous SNe (SLSNe) prefer low mass (metal poor) galaxies with high specific star formation rates \citep{kelly2012core,schulze2021palomar}. On local scales, SESNe strongly trace H$\alpha$ (a tracer of ongoing star formation), while Type II SNe tend to show larger spreads in local environment properties \cite{anderson2012progenitor}. In contrast the Type IIn SNe, which show signs of interaction between the SN blast wave and dense, pre-existing circumstellar material, show highly heterogeneous local environments \citep{ransome2022h}. The rarer Ca-rich SNe of yet unknown progenitor origin show strong preference for high offsets in their host galaxies, although they are not preferentially found in star-forming hosts \citep{kasliwal2012calcium, dong2022physical}. These correlations are clues toward the underlying progenitor populations for each of these SN classes. Here, however, we utilize the correlations between global host properties and transients as a tool to help classify the underlying physics of the SNe.

Classifiers that utilize galaxy information for SN identification have been explored in the literature, although they have been primarily limited to binary classification tasks. \cite{foley2013classifying} introduced the first galaxy-based SN classifier based on a Naive Bayes architecture with the LOSS SN sample. In a similar vein, \cite{baldeschi2020star} showed that galaxy morphology and star formation could be used to increase the purity of thermonuclear and CCSN samples compared to randomly guessing. \cite{gagliano2021ghost} introduced a random forest classifier based on host galaxy properties from the \texttt{GHOST} sample of $\simeq16,000$ SNe associated with host galaxies, finding that such a method could perform thermonuclear vs CCSNe classification with $\simeq68$\% accuracy. \cite{gomez2020fleet} presented one of the first host-aware classifiers which extend beyond Ia/CCSNe classification, focusing instead on Type I SLSNe vs non-SLSN classification from a combination of host and light curve properties; a classifier based on host galaxy properties alone is not explored in that work. \cite{2022arXiv220902784K} presented a hierarchical SN classifier based solely on optical and (near) infrared host galaxy photometry from the THEx catalog \citep{qin2022linking}. They use a likelihood-based approach reminiscent of a Naive Bayes classifier, using a series of binary classifications to distinguish 11 SN classes (in addition to tidal disruption events). They were able to classify Type Ia subtypes and Type II SNe with a purity statistically above random guessing. Most recently, \cite{gagliano2023first} presented a neural network-based classifier for Zwicky Transient Facility (ZTF) Bright Transient Survey data which utilizes the light curve and host galaxy information for Type Ia, II and Ibc SN classification, achieving an accuracy of $\simeq82$\% within 3-days (observer-frame) of discovery.

Furthermore, the hierarchical nature of SN taxonomy is rarely used in photometric classification. \cite{sanchez2021alert} presented a broad hierarchical classifier for ZTF data. There, multiple``flat" (i.e., those lacking a hierarchical structure) random forest algorithms are trained at each ``level" of the hierarchical classifier. For example, all SNe are classified as transient phenomena, then a second flat classifier is trained to classify them as one of four SN types. \cite{2022arXiv220902784K} consider the hierarchical nature of SNe in the sense of a prior distribution, which is then used in the classifier's objective function. For example, the rate of Type Ia-91bg events are included when calculating the rate of Type Ia SNe, which is then used to define a Type Ia prior probability within the objective function. However, this prior information is not used in their primary analysis.

Here, we present a broad, hierarchical classifier based on host galaxy properties alone and explore how galaxy properties can improve our transient classifications. We specifically test how simple, measurable properties of the host impact our classification accuracy both with and without redshift information. The paper is organized as follows. In Section~\ref{sec:data}, we review the data used in this work and the features selected for our SN classifiers. In Section~\ref{sec:methods} we present the weighted hierarchical cross-entropy score to more naturally account for the hierarchical nature of SN classification. We additionally discuss the architecture of our neural network-based classifier used here for classification. In Section~\ref{sec:results}, we discuss the results of our classifiers for different subsets of SN classes, feature sets and the in/exclusion of redshift information. In Section~\ref{sec:ps1}, we present new classifications of SNe within the Pan-STARRs Medium Deep Survey sample. We conclude in Section~\ref{sec:conclusions}. Throughout this work, we assume a flat $\Lambda$CDM cosmology with H$_0 = 67.8 $ km s$^{-1}$ Mpc$^{-1}$, $\Omega_\mathrm{m} = 0.308$\citep{collaboration2020planck}.

\section{Data \& Feature Selection}\label{sec:data}

\begin{figure*}[t]
\centering
\includegraphics[width=0.8\textwidth]{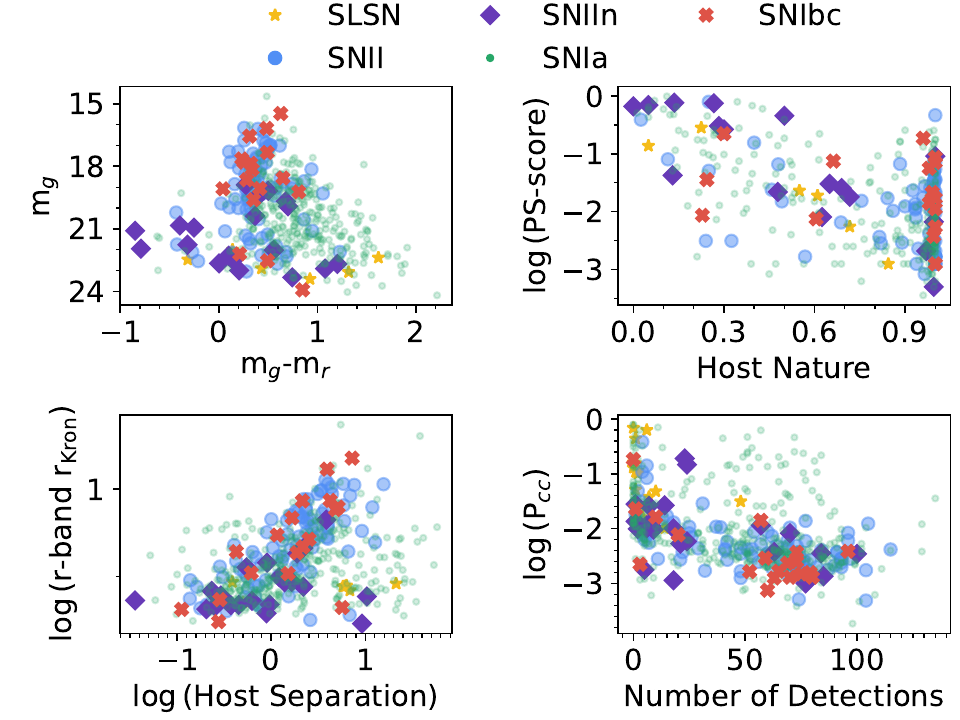}
\caption{Scatter plot of various host galaxy features to classify SNe in this work (see Sec.~\ref{sec:data} for feature definitions). SNe are \textit{spectroscopically} labelled as part of PS1 MDS, and for features where the host galaxy features were not observed, they are inferred via K-means imputation. Even in these simple feature spaces, clear clustering is seen. \label{fig:hostscatter}}
\end{figure*}

We aim to understand \textit{if} and \textit{when} host galaxy features can aid in SN photometric classification. Therefore, we require a SN sample for which both a light curve \textit{only} classifier and a host galaxy classifier are both available. While there now exist two large catalogs of host galaxies with associated SNe, GHOST \cite{gagliano2021ghost} and THEx \cite{qin2022linking}, we instead opt to use a uniform survey in order to directly compare the light curve vs host features. For this reason, we utilize the Pan-STARRS Medium Deep Survey catalog of SN-like light curves published by \cite{villar2020superraenn} and \cite{hosseinzadeh2020photometric}.  In total, our sample includes 557 spectroscopically identified SNe, in five classes: Type Ia SNe (404 objects), Type II SNe (93 objects), Type IIn SNe (24 objects), Type Ib/c SNe (19 objects), and Type I SLSNe (17 objects). Our Pan-STARRS sample has the additional benefit of being a close analogue to Vera C. Rubin data in terms of cadence, filter selection and depth, offering a realistic view of how our algorithm will perform on the LSST datastream.

We use the full set of light curve features available from \texttt{SuperRAENN} (44 in total; \citealt{villar2020superraenn}), which includes:

\begin{enumerate}
    \item [1-8:] Non-linear features extracted from an autoencoder, trained on the PS1-MDS dataset. The correlations between these data-driven features and ``typical" observational features are explored in more detail in \cite{villar2020superraenn}.
    \item [9-21:] The $griz$ rise times, calculated using 1, 2, and 3 magnitudes below peak. Note that these are calculated using light curves which have been interpolated via a 2D Gaussian Process. These are, when redshift is utilized, calculated in the rest frame of the SN (using only the cosmological k-corrections).
    \item [22-32:] The $griz$ decline times, calculated for 1, 2 and 3 magnitudes below peak.
    \item [33-36]: The $griz$ peak magnitude. When redshift is utilized, these are absolute magnitudes.
    \item [37-40:] The median slope measured in $griz$ between 10 and 30 days post-peak in the observer frame. Note that these features in particular help distinguish between Type II and Type Ibc SNe.
    \item [41-44:] The integrals of the interpolated $griz$ light curves.
\end{enumerate}

To associate SNe with host galaxies and extract galaxy features, we use the ``Finding Luminous and Exotic Extragalactic Transients" (FLEET) pipeline \citep{gomez2020fleet,gomez2023first}. FLEET is a classification methodology designed to classify rare, extragalactic transients (SLSNe and tidal disruption events) using a combination of host galaxy and light curve information.  FLEET queries a $1'$ region of the Pan-STARRS $3\pi$ survey around a given transient to identify the most likely host galaxy using the follow algorithm. First, a probability of being a galaxy (as opposed to a star) is assigned to every object in the field, where 0 means most like a star, and 1 means most likely a galaxy. This probability is estimated using a custom k-nearest neighbors algorithm trained on data from the Canada-France-Hawaii Telescope Legacy Survey (CFHTLS), which has complete star/galaxy labels down to $\approx 26$ mag \citep{hudelot2012vizier}. Then, the probability of chance coincidence $P_{\rm cc}$ of every galaxy in the search region is calculated using the \cite{bloom2002observed} method described in \cite{berger2010short}. More precisely, we follow Equation 2 of \citealt{gomez2020fleet}:
\begin{align}
        P_\mathrm{CC}  &= 1 - e^{-\pi (d^2 + 4R^2)\Sigma(\leq m)}\\
        \Sigma(\leq m) &= \frac{10^{0.33(m-24)-2.44}}{0.33\ln(10)}
\end{align}
where $d$ is the angular separation between the transient and its host in arcseconds; $R$ is the half-light radius of the host also in arcseconds; and $m$ is the $i$-band magnitude of the host (or the $r$-band is $i$-band is not measured). We select the galaxy with the lowest $P_{\rm cc}$ as the host galaxy of the transient, or (for SNe in which $P_{\rm cc}>0.1$) consider the transient to be ``hostless". Hostless transients are retained in the dataset, as these are often associated with rare CCSN types (e.g., SLSNe and Type IIn). The selection of $P_{\rm cc}>0.1$ is somewhat arbitrary, but found to be a reasonable threshold in \citealt{gomez2020fleet}.

Many host galaxy features are available via FLEET, including multi-survey observables (e.g., magnitudes), derived properties (e.g., photo-z), and inferred properties more directly related to the transient (e.g., offset). We restrict ourselves to data from the Pan-STARRS $3\pi$ survey, in order to minimize missing features in our dataset. We use the following features from the pipeline:

\begin{itemize}
    \item [1-5:] $g$-,  $r$- $i$-, $z$- and $y$-band Kron magnitudes, estimated using \texttt{sextractor}.
    \item [6-8:] $g$-, $r$- and $i$-band Kron radii, estimated using \texttt{sextractor}.
   \item [9:] Host galaxy separation from the transient in arcseconds. If the transient is deemed ``hostless", the offset value is set to zero.
    \item [10:] Point source score, a measurement of probability of the object being an extended object (score of 0) or a point source (score of 1). This is a property within the PS1 $3\pi$ Survey.
    \item [11:] Number of detections in all bands from the independent images used to generate the stacked image.
    \item [12:] Host galaxy half-light radius in $i$-band or in $r$-band if $i$-band is not available.
    \item [13:] Host galaxy nature, a measurement of probability of the object being a galaxy (score of 1) or a star (score of 0). This is a custom function within FLEET.
    \item [14:] Probability of chance coincidence, as calculated by FLEET.
\end{itemize}

All magnitudes are corrected for Milky Way extinction. In some of our pipelines, we additionally include the redshift (spectroscopically measured from the host or transient) as a feature.


Although we attempt to minimize missing data, a small fraction of events are missing some observational features. Most often (in 4 cases of the 557 objects in the spectroscopic dataset), the point source score and uncertainties on the measured magnitudes are missing due to a transient being ``hostless". We use a K-means imputation method (\texttt{KNNImputer} in \texttt{scikit-learn}) to fill in the missing data. This method utilizes the information from the $K=3$ neighbors in the 13-dimensional galaxy feature space, and fills in missing values with an average from these neighbors. This imputation method works as expected with no ad-hoc corrections: for hosts which are not detected due to missing data (e.g., they were never observed in $z$-band), the nearest neighbor provides a reasonable estimate of the missing properties. For hosts near the survey limit, the method naturally fills in the missing data with dim apparent magnitudes and small observed radii. In testing, we find that changing the value of $k$ has minimal impact on results. Finally, we normalize our data such that each feature values between 0 and 1 (\texttt{MinMaxScaler} in \texttt{scikit-learn}). 

We note that there are clear correlations between SN type and host galaxy properties, visible without the aid of a specialized classifier. We show a number of representative feature spaces in Fig.~\ref{fig:hostscatter}. Brighter objects (SLSNe, Type IIn SNe) tend to occur in dimmer galaxies. This is likely an observational bias of PS1-MDS, as our SLSNe and Type IIn SNe sample skew to higher redshifts (see Figure 1 of \citealt{villar2019supernova}). The color of these galaxies also clearly strongly correlates with type. For any given magnitude, CC SNe (especially Type II SNe) are more likely to occur in bluer galaxies (i.e., those with ongoing star formation). 

\section{Classification Methods}\label{sec:methods}

\subsection{A Novel Hierarchical Loss Function}

\begin{figure*}[t]
\includegraphics[width=0.95\textwidth,trim={0 8cm 5cm 0},clip]{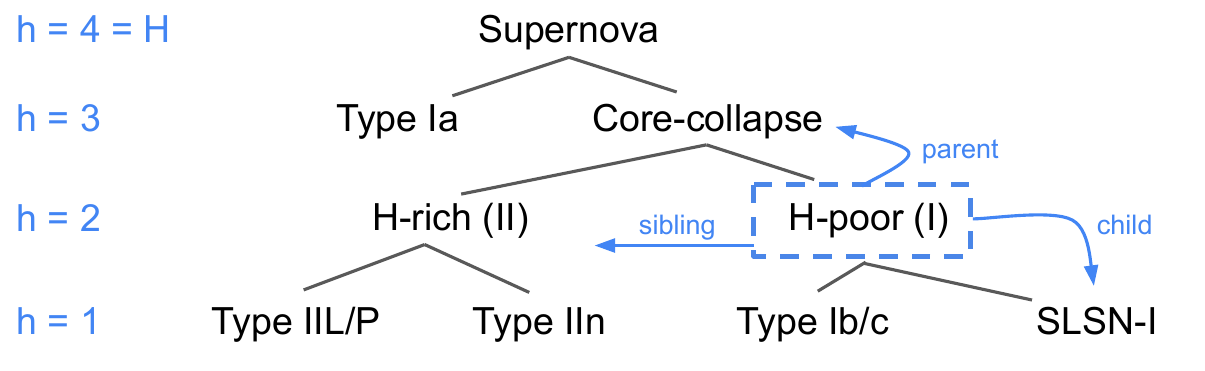}
\caption{Hierarchical graph structure used in this work. Blue text is meant as a guide to the various components of a generic tree. \label{fig:diagram}}
\end{figure*}


We explicitly include the hierarchical nature of SN taxonomy within our analysis via a weighted hierarchical cross-entropy (WHXE) objective function (also presented in \citealt{villar2023hierarchical} for SNe and variable stars). \cite{bertinetto2020making} originally introduced the HXE to classify images with a similar hierarchical taxonomy. They compare the performance of HXE to standard cross-entropy, finding that the two perform similarly in overall accuracy but that HXE ``makes better mistakes", i.e., the HXE enforces a graph structure that encourages the classifier to place objects in the correct broader category. In our SN context, the WHXE allows us to (1) train a multi-layered classifier that can easily perform binary- and general-classification and (2) encourage transients to be misclassified within their broader class (e.g., CCSNe are less likely to be classified as Type Ia SNe). In this Section, we define the graph structure used for our SN classification and WHXE in detail. 

The graphical nature of our classification hierarchy is shown in Fig.~\ref{fig:diagram}. Given the small dataset size, we have not subdivided Type Ia, Type IIP/L and Type Ib/c SNe into additional subclasses. We emphasize, however, that this is not the \textit{only} choice for a classification graph. For example, some bright Type IIn SNe share similar hosts to SLSNe. One could include galaxy information in the graph structure (e.g., preferring to classify bright transients in low-luminosity galaxies together), although the classification bias would be less physically motivated. Alternatively, one could follow the ``classic" classification schema, first distinguishing Type I vs Type II SNe (rather than thermonuclear vs core-collapse). This may aid in isolating hydrogen-rich transients, but is not explored in this work.

We next describe the WHXE objective function, largely following the notation of\ \cite{villar2023hierarchical}. The classic categorical cross-entropy, most often used in classification tasks, is defined as:
\begin{equation}\label{eqn:ce-loss}
    \mathcal{L}= - \sum_i^C t_i \log(p(c_i)),
\end{equation}
where $C$ is the total number of classes, $c_i$ is a specific SN class, and $t_i$ is an indicator variable for the \textit{true} SN class. This indicator function means that the cross-entropy score \textit{only} rewards the classifier based on assigning a high likelihood to the true class. There is no reward structure for how probabilities are distributed over the incorrect classes.

We contrast this to the WHXE, which utilizes the hierarchical graph structure of the SN taxonomy by factorizing each class by its leaf nodes. We represent the SN class ``height" on the tree as $c_i^{(h)}$. The root (the ``SN" parent class) is represented as $c_i^{(H)}$ (where $H=4$ is the height of the tree), and a lowest leaf (e.g., the Type Ib/c class) is represented as $c_i^{(1)}$. For any given class $c_i$ with height $h'$, the classification probability can be written as:

\begin{equation}
    p(c_i) = \prod^{H-1}_{h=h'}p(c_j^{(h)}|c_k^{(h+1)})
\end{equation}
where $c_j$ and $c_k$ are ancestor nodes of class $c_i$ (i.e., nodes on the path from the leaf to the root). Here, we are assuming that the probability of being a SN (the root node) is equal to one. The conditional probabilities can be written more explicitly as

\begin{equation}
p(c_j^{(h)}|c_k^{(h+1)}) =
\frac{p(c_j^{(h)})}
{p(c_k^{(h+1)})},
\end{equation}

where $c_j^{(h)}$ is a child node of $c_k^{(h+1)}$. Because our classifier predicts probabilities for the leaf classes, the probability of any internal node is obtained by summing the probabilities of all descendant leaf classes:

\begin{equation}
p(c)=\sum_{l\in\mathrm{Leaves}(c)}p(l).
\end{equation}

As an example, the probability that a SN is Type Ib/c given that it is H-poor is

\begin{equation}
p(\mathrm{Ib/c}|\mathrm{H\mbox{-}poor})
=
\frac{p(\mathrm{Ib/c})}
{p(\mathrm{H\mbox{-}poor})},
\end{equation}

where p(H-poor) is the sum of the probabilities of all H-poor descendant classes.

Analogous to Eqn.\ref{eqn:ce-loss}, we can define the weighted hierarchical cross-entropy (WHXE) as:
\begin{equation}
    \mathcal{L}_\mathrm{WHXE}(c_i^{(h)})=-\sum_{h=h'}^{H-1}W(c^{(h)})\lambda(c^{(h)})\log p(c_j^{(h)}|c_k^{(h+1)}),
\end{equation}
where $W(c^{(h)})$ weights each SN class, and $\lambda(c^{(h)})$ is a weighting which emphasizes each level of the taxonomy tree (as a function of class height, $h$). For $W(c^{(h)})$, we choose to weight each SN by $N_\mathrm{All} / (N_\mathrm{Labels}\times N_\mathrm{N_c})$, where $N_\mathrm{All}$ is the total number of SNe in our training set, $N_\mathrm{Labels}$ is the number of SN classes (all nodes), and $N_c$ is the number of SNe of type $c$. \cite{bertinetto2020making} suggest the following form for $\lambda(c^{(h)})$:
\begin{equation}
    \lambda(c^{(h)}) = \exp(-\alpha h),
\end{equation}
where $\alpha$ is a free parameter. Larger values of $\alpha$ weight the top of the hierarchy more strongly (i.e., Ia vs CC classification); lower values of $\alpha$ weight each level of the hierarchy equally, emphasizing fine-grained classifications; $\alpha$ is therefore a hyperparameter of our model which we optimize. 

Based on our graph in Figure~\ref{fig:diagram}, our objective function requires nine outputs from our classifier, reflecting a probability at each node of the graph. The top node (the ``SN" designation) is always equal to one. A softmax function is then applied to each branch of the tree such that the neural network output can be interpreted as a conditional probability. For example, one output represents the probability\footnote{Note: these are not ``true" probabilities as they are not properly calibrated. Instead, they share qualitative similarities to a probability vector: they contain non-negative values which sum to one.} of being H-rich and another represents the probability of being H-poor. As these are a pair of siblings, these outputs will be re-normalized such that their sum is equal to one. The height of each output is also tracked, with an appropriate weighting (a function of $\alpha$) applied. A working \texttt{PyTorch} implementation of the WHXE is presented in \cite{villar2023hierarchical} and available via GitHub\footnote{\url{https://github.com/VTDA-Group/hxe-for-tda/}}.

\subsection{Multi-layer Perceptron Classifier}

Throughout this work, we use a fully connected multi-layer perceptron (MLP, a simple neural network) to classify the SN subclasses. The MLP transforms an input feature vector into an output probability vector which optimizes the WHXE objective function. Between these are a series of ``hidden" layers with optimizable weights and nonlinear activation functions. Here, we use the standard Rectified Linear Unit. We optimize the MLP using the standard Adam optimizer \citep{kingma2014adam}, a momentum-based gradient descent algorithm. Our model and training procedure is built in \texttt{Pytorch}.

In total, our hyperparameters of the model are: (1) the hierarchy weighting, $\alpha\in\{0.0,0.1,1.0,3.0\}$; (2) the learning rate, $\beta\in\{0.001, 0.005, 0.01, 0.05, 0.1\}$; (3) the batch size $\in\{16, 32, 64, 128, 256\}$; and (4) the number of neurons per layer, a number from the set $\in\{3, 5, 10\}$. Using each combination of hyperparameters, we train a MLP for 300 epochs with early stopping and select the optimal set for each feature set. We train six unique classifiers using different light curve / host features (see next Section for details). However, the best hyperparameter values (as determined by tracking the F1-score) do not vary greatly. Typically, the best hierarchy weighting is $\alpha\simeq1$, the  learning rate on the higher end $\beta\simeq0.05$, the batch size is on the higher end ($128-256$) and the number of neurons is on the higher end ($5-10$). Our model takes minutes on a single CPU to complete training. We note that we also attempted to use synthetic minority oversampling techniques to re-weight our training set (as opposed to including the class weights in our objective function), finding somewhat worse results.

\begin{deluxetable*}{lccccccc}
\tabletypesize{\footnotesize}
\tablecolumns{8}
\tablecaption{ Classification Performance \label{table:results1}}
\tablehead{
\colhead{} & \colhead{} & \colhead{Gal. w/o z} & \colhead{Gal. w/ z} &  \colhead{LC w/o z} &\colhead{LC w/ z} & \colhead{Gal.+LC w/o z} & \colhead{Gal. + LC w/ z} }
 \startdata\hline
 \multicolumn{8}{c}{Five-way Classification}\\\hline
SLSN & F1 & 0.19 (0.03) &  0.63 (0.11) &  0.23 (0.02) &  \textbf{0.77 (0.05)} &  0.29 (0.05) &  0.73 (0.05)\\
 &  Purity & 0.12 (0.02) &  0.50 (0.14) &  0.20 (0.02) &  \textbf{0.73 (0.07)} &  0.25 (0.06) &  \textbf{0.73 (0.06)}\\
 &  Completeness & 0.47 (0.06) &  \textbf{0.83 (0.09)} &  0.27 (0.04) &  0.80 (0.05) &  0.33 (0.09) &  0.73 (0.08)\\
II & F1 & 0.38 (0.02) &  0.46 (0.06) &  0.52 (0.03) &  0.71 (0.03) &  0.56 (0.04) &  \textbf{0.73 (0.02)}\\
 &  Purity & 0.39 (0.03) &  0.41 (0.04) &  0.55 (0.05) &  \textbf{0.79 (0.04)} &  0.62 (0.04) &  \textbf{0.79 (0.03)}\\
 &  Completeness & 0.36 (0.03) &  0.53 (0.14) &  0.51 (0.05) &  0.65 (0.03) &  0.51 (0.05) &  \textbf{0.67 (0.02)}\\
IIn &F1 & 0.16 (0.02) &  0.16 (0.03) &  0.34 (0.04) &  0.38 (0.06) &  0.37 (0.04) &  \textbf{0.41 (0.04)}\\
 &  Purity & 0.10 (0.02) &  0.11 (0.02) &  0.29 (0.05) &  0.32 (0.08) &  0.29 (0.04) &  \textbf{0.41 (0.06)}\\
 &  Completeness & 0.33 (0.05) &  0.27 (0.10) &  0.42 (0.09) &  0.46 (0.06) &  \textbf{0.50 (0.07)} &  0.42 (0.05)\\
Ia &F1 & 0.70 (0.02) &  0.73 (0.04) &  0.88 (0.01) &  \textbf{0.94 (0.02)} &  0.91 (0.01) &  \textbf{0.94 (0.01)}\\
 &  Purity & 0.86 (0.01) &  0.91 (0.01) &  0.93 (0.01) &  \textbf{0.95 (0.01)} &  0.94 (0.01) &  0.93 (0.01)\\
 &  Completeness & 0.60 (0.03) &  0.61 (0.06) &  0.84 (0.02) &  0.93 (0.04) &  0.89 (0.01) &  \textbf{0.95 (0.01)}\\
Ibc &F1 & 0.16 (0.03) &  0.13 (0.05) &  0.27 (0.04) &  0.31 (0.04) &  0.29 (0.03) &  \textbf{0.36 (0.04)}\\
 &  Purity & 0.12 (0.03) &  0.09 (0.04) &  0.19 (0.04) &  0.24 (0.04) &  0.22 (0.03) &  \textbf{0.32 (0.04)}\\
 &  Completeness & 0.26 (0.07) &  0.29 (0.19) &  \textbf{0.47 (0.06)} &  \textbf{0.47 (0.08)} &  0.42 (0.08) &  0.42 (0.06)\\
Average &F1 & 0.36 (0.01) &  0.45 (0.03) &  0.46 (0.01) &  0.63 (0.02) &  0.49 (0.02) &  \textbf{0.64 (0.02)}\\
 & Purity & 0.32 (0.01) &  0.41 (0.03) &  0.43 (0.02) &  0.61 (0.02) &  0.46 (0.02) &  \textbf{0.64 (0.02)}\\
 &  Completeness & 0.40 (0.02) &  0.51 (0.06) &  0.50 (0.03) &  \textbf{0.66 (0.02)} &  0.53 (0.03) &  0.64 (0.02)\\
Weighted Avg. &F1 & 0.60 (0.02) &  0.65 (0.03) &  0.76 (0.01) &  \textbf{0.85 (0.02)} &  0.79 (0.01) &  \textbf{0.85 (0.01)}\\
 &  Purity & 0.69 (0.01) &  0.75 (0.01) &  0.79 (0.01) &  \textbf{0.86 (0.01)} &  0.81 (0.01) &  0.85 (0.01)\\
 &  Completeness & 0.53 (0.03) &  0.58 (0.05) &  0.73 (0.02) &  0.84 (0.03) &  0.77 (0.01) &  \textbf{0.85 (0.01)}\\\hline
 \multicolumn{8}{c}{Three-way Classification}\\\hline
II & F1 & 0.44 (0.02) &  0.48 (0.04) &  0.66 (0.02) &  \textbf{0.75 (0.04)} &  0.71 (0.02) &  \textbf{0.75 (0.02)}\\
 &   Purity & 0.36 (0.02) &  0.37 (0.04) &  0.65 (0.03) &  \textbf{0.79 (0.07)} &  0.72 (0.02) &  \textbf{0.79 (0.03)}\\
 &   Completeness & 0.58 (0.03) &  0.70 (0.10) &  0.68 (0.04) &  \textbf{0.72 (0.03)} &  0.70 (0.03) &  \textbf{0.72 (0.02)}\\
Ia &F1 & 0.68 (0.03) &  0.70 (0.05) &  0.88 (0.01) &  \textbf{0.94 (0.02)} &  0.91 (0.01) &  \textbf{0.94 (0.01)}\\
 &   Purity & 0.88 (0.02) &  0.92 (0.01) &  0.93 (0.01) &  \textbf{0.95 (0.01)} &  0.94 (0.01) &  0.93 (0.01)\\
 &   Completeness & 0.56 (0.04) &  0.57 (0.06) &  0.83 (0.02) &  0.92 (0.04) &  0.88 (0.01) &  \textbf{0.95 (0.01)}\\
SE &F1 & 0.20 (0.02) &  0.31 (0.06) &  0.30 (0.03) &  0.48 (0.04) &  0.38 (0.03) &  \textbf{0.52 (0.03)}\\
 &   Purity & 0.14 (0.02) &  0.22 (0.06) &  0.22 (0.03) &  0.38 (0.04) &  0.29 (0.03) &  \textbf{0.47 (0.04)}\\
 &   Completeness & 0.41 (0.06) &  0.51 (0.10) &  0.47 (0.05) &  \textbf{0.65 (0.05)} &  0.53 (0.06) &  0.59 (0.05)\\
Average &F1 & 0.49 (0.01) &  0.54 (0.03) &  0.63 (0.01) &  0.73 (0.02) &  0.68 (0.01) &  \textbf{0.74 (0.01)}\\
 &  Purity & 0.46 (0.01) &  0.50 (0.02) &  0.60 (0.01) &  0.71 (0.03) &  0.65 (0.01) &  \textbf{0.73 (0.02)}\\
 &  Completeness & 0.52 (0.03) &  0.59 (0.05) &  0.66 (0.02) &  \textbf{0.76 (0.03)} &  0.70 (0.02) &  0.75 (0.02)\\
Weighted Avg. &F1 & 0.62 (0.02) &  0.66 (0.03) &  0.80 (0.01) &  \textbf{0.87 (0.02)} &  0.83 (0.01) &  \textbf{0.87 (0.01)}\\
 &  Purity & 0.72 (0.01) &  0.75 (0.01) &  0.82 (0.01) &  \textbf{0.88 (0.02)} &  0.85 (0.01) &  0.87 (0.01)\\
 &  Completeness & 0.55 (0.03) &  0.59 (0.05) &  0.78 (0.02) &  0.86 (0.03) &  0.82 (0.01) &  \textbf{0.88 (0.01)}\\\hline
 \multicolumn{8}{c}{Binary Classification}\\\hline
Ia &F1 & 0.54 (0.02) &  0.62 (0.03) &  0.77 (0.02) &  \textbf{0.85 (0.03)} &  0.81 (0.01) &  \textbf{0.85 (0.02)}\\
 &   Purity & 0.44 (0.02) &  0.49 (0.03) &  0.73 (0.03) &  0.85 (0.06) &  0.78 (0.01) &  \textbf{0.88 (0.02)}\\
 &   Completeness & 0.68 (0.04) &  0.84 (0.04) &  0.82 (0.02) &  \textbf{0.86 (0.02)} &  0.83 (0.01) &  0.82 (0.02)\\
CC &F1 & 0.74 (0.02) &  0.76 (0.04) &  0.90 (0.01) &  \textbf{0.94 (0.02)} &  0.92 (0.01) &  \textbf{0.94 (0.01)}\\
 &   Purity & 0.84 (0.01) &  0.91 (0.01) &  0.93 (0.01) &  \textbf{0.94 (0.01)} &  0.93 (0.01) &  0.93 (0.01)\\
 &   Completeness & 0.65 (0.04) &  0.65 (0.05) &  0.87 (0.02) &  0.94 (0.04) &  0.91 (0.01) &  \textbf{0.96 (0.01)}\\
Average& F1 & 0.66 (0.01) &  0.72 (0.02) &  0.84 (0.01) &  \textbf{0.90 (0.02)} &  0.86 (0.01) &  \textbf{0.90 (0.01)}\\
 &  Purity & 0.64 (0.01) &  0.70 (0.02) &  0.83 (0.01) &  0.90 (0.03) &  0.86 (0.01) &  \textbf{0.91 (0.01)}\\
 &  Completeness & 0.67 (0.03) &  0.74 (0.03) &  0.85 (0.01) &  \textbf{0.90 (0.02)} &  0.87 (0.01) &  0.89 (0.01)\\
Weighted Avg. &F1 & 0.69 (0.02) &  0.74 (0.02) &  0.86 (0.01) &  \textbf{0.92 (0.02)} &  0.89 (0.01) &  \textbf{0.92 (0.01)}\\
 &  Purity & 0.73 (0.01) &  0.79 (0.01) &  0.87 (0.01) &  \textbf{0.92 (0.02)} &  0.89 (0.01) &  \textbf{0.92 (0.01)}\\
 &  Completeness & 0.66 (0.03) &  0.70 (0.04) &  0.86 (0.01) &  \textbf{0.92 (0.03)} &  0.88 (0.01) &  \textbf{0.92 (0.01)}\\
\enddata
 \tablecomments{Classification performance (quantified by accuracy, purity and completeness) using a 5-way, 3-way and 2-way split. Optimal feature sets are bolded for each category. }
\end{deluxetable*}

\section{Classification Results \& Discussion}\label{sec:results}

Our goal is to understand how contextual host galaxy information improves classification performance of SNe both with and without redshift information. Therefore, we train and compare six classifiers in total: (1) one which uses solely the observer-frame host galaxy information; (2) one which uses observer-frame host galaxy information and redshift; (3) one which uses solely the observer-frame light curve information; (4) one which uses the rest-frame light curve and redshift information; (5) one which uses the galaxy information and the observer-frame light curve information and (6) one which uses the galaxy, rest-frame light curve and redshift information.

We track the purity, completeness and F1-score of each classifier, defined as:
\begin{align}
    \mathrm{Purity} &= \frac{\mathrm{TP}}{\mathrm{TP}+\mathrm{FP}}\\
        \mathrm{Completeness} &= \frac{\mathrm{TP}}{\mathrm{TP}+\mathrm{TN}}\\
        \mathrm{F1} &= \frac{2\times(\mathrm{Purity}\times\mathrm{Competeness})}{\mathrm{Purity}+\mathrm{Completeness}}.
\end{align}
TP is the ``true positive" rate or the fraction of SNe within a given class correctly identified as belonging to said class. TN is the true negative rate or the fraction of SNe \textit{not} in a given class and correctly identified as \textit{not} being a member of said class. Finally, FP is the false positive rate or the fraction of SNe which are identified as belonging to a given class but in fact are \textit{not} of that class. The F1-score is the harmonic mean of the purity and completeness and a commonly used metric for evaluating classifiers.

We report the purity, completeness and F1-score for each SN class, as well as the class-averaged and the ``weighted'' versions of these metrics. Here, the ``weighted" averages re-weight each class to represent the total number of objects in each class. In this case, the statistic is dominated by the majority class, Type Ia SNe. Note that we do not use accuracy to evaluate each classifier, which can be a particularly poor metric of success for highly imbalanced training sets; however, when appropriate (i.e., when comparing to other works), we report accuracy of specific models.

We test our classifier on three classification tasks: (1) five-way classification (SLSN, Type II, Type IIn, Type Ia and Type Ibc); (2) three-way classification (Type II, Type Ia, and stripped-envelope); and (3) binary classification (core-collapse vs Type Ia). As described above, the classifier does not need to be retrained for each of these tasks; instead, we calculate the conditional probabilities of each, assigning the final label of each object as the category with the highest conditional probability, e.g. for Type IIn SNe: 

\begin{equation}
    p(\mathrm{IIn})=p(\mathrm{CC})P(\mathrm{H{\text -}rich}|\mathrm{CC})P(\mathrm{IIn}|\mathrm{H{\text -}rich})
\end{equation}

\begin{figure*}[t]
\includegraphics[width=\textwidth, trim=0 20 0 0, clip]{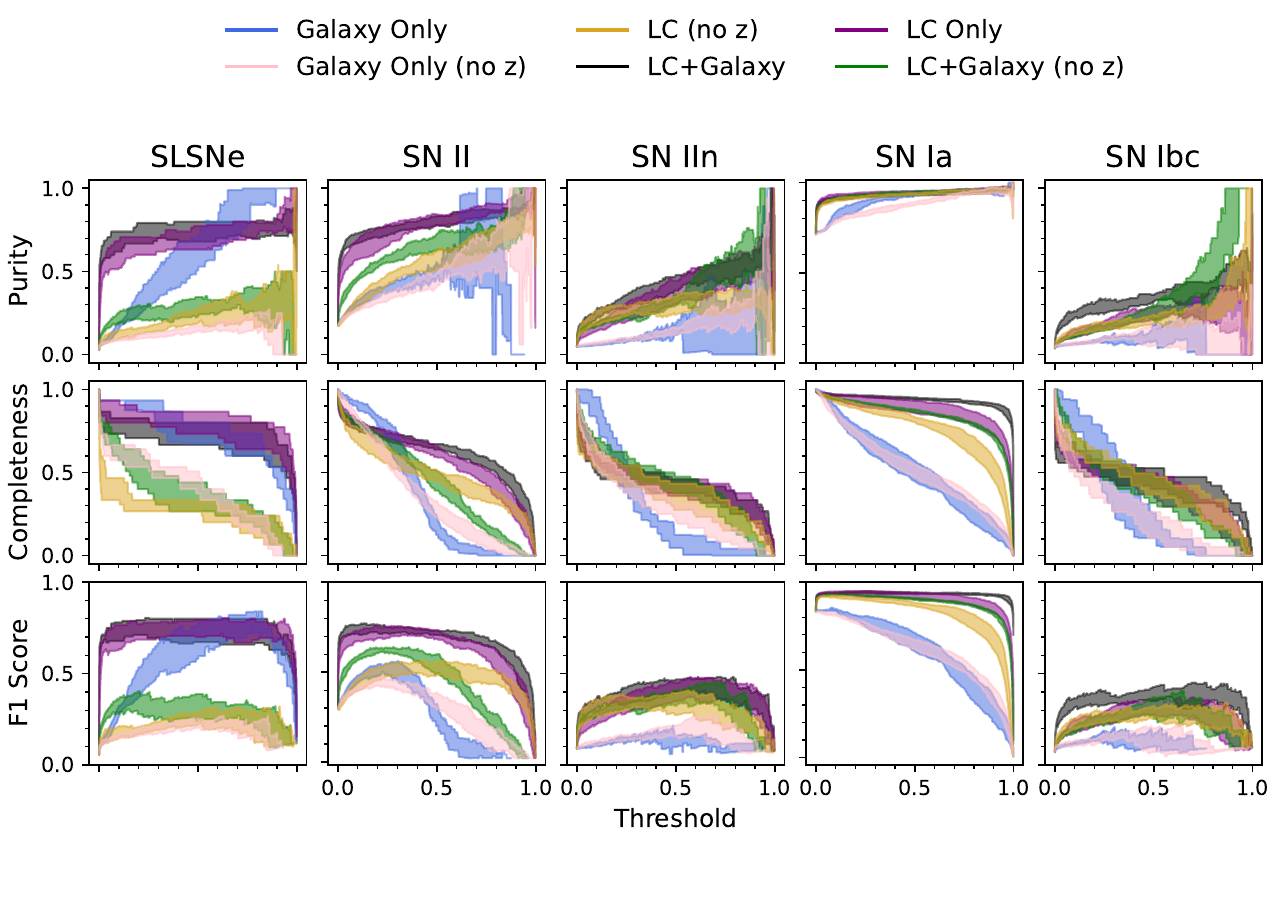}
\caption{Purity (\textit{top row}), completeness (\textit{middle row})  and F1-score (\textit{bottom row}) achieved with the various feature sets and confidence thresholds as a function of SN type (\textit{columns}).  Shaded regions represent $1\sigma$ uncertainties, computed using the same classifier with different random model initializations during training. \label{fig:purcom}}
\end{figure*}

In total, we compare 18 combinations of feature sets and output classes. Our results are fully summarized in Table~\ref{table:results1} and visualized summarized in Figure~\ref{fig:purcom}. Throughout this section, we will exclude uncertainties (which have been calculated using ten random seeds for each MLP) for purity, completeness and  F1-scores. However, these values are listed within Table~\ref{table:results1} and are typically $\lesssim0.05$. 

\begin{figure*}[t]
    \centering
    \parbox{0.45\textwidth}{
        \centering
        \includegraphics[width=\linewidth]{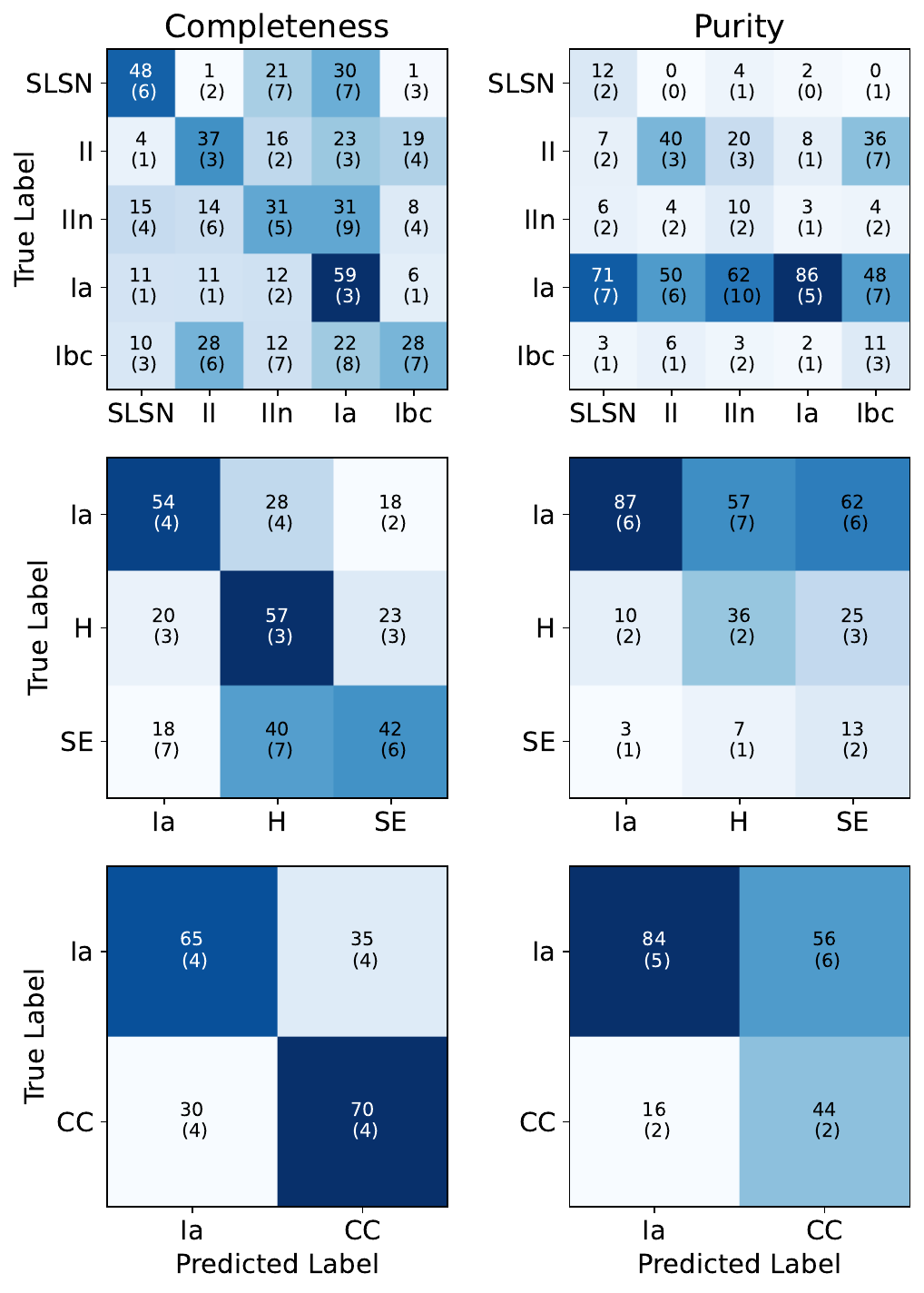}\\
        (a) Galaxy Features, Excluding Redshift
    }
    \qquad
    \parbox{0.45\textwidth}{
        \centering
        \includegraphics[width=\linewidth]{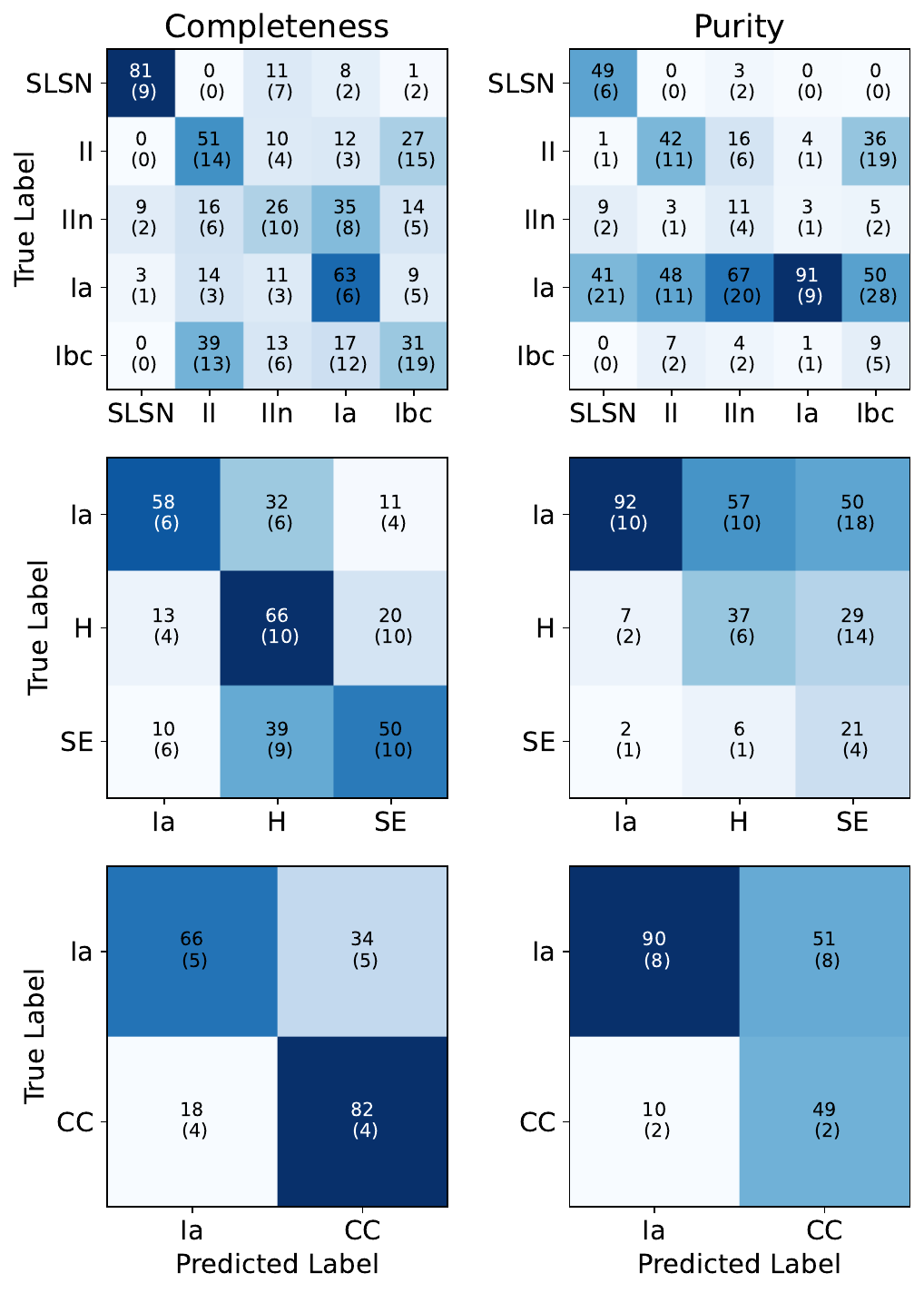}\\
        (b) Galaxy Features, Including Redshift
    }
    \caption{Confusion matrices for the classifier using \textit{only} host galaxy information without (\textit{left}) and with (\textit{right}) redshift information. Numbers are overall percent, while uncertainties are given in parentheses (e.g., the completeness of SLSNe in the five-way classifier is $0.48\pm0.06$). The classifier reaches state-of-the-art results for Type Ia vs CC classification, but fails to achieve high accuracy in the five-way class split.}
    \label{fig:conf-gal-all}
\end{figure*}

\begin{figure*}[t]
    \centering
    \parbox{0.45\textwidth}{
        \centering
        \includegraphics[width=\linewidth]{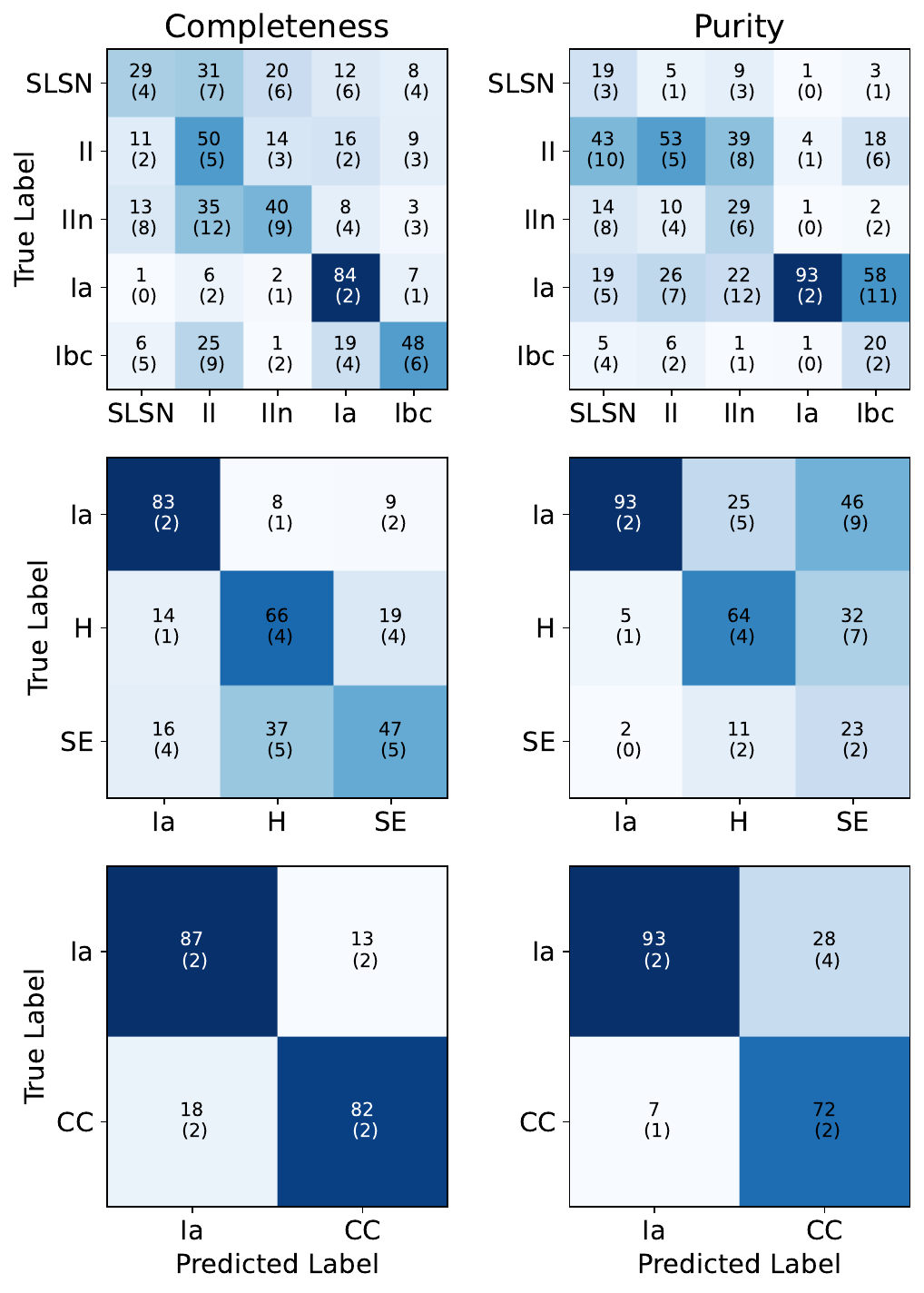}\\
        (a) Light Curve Features, Excluding Redshift
    }
    \qquad
    \parbox{0.45\textwidth}{
        \centering
        \includegraphics[width=\linewidth]{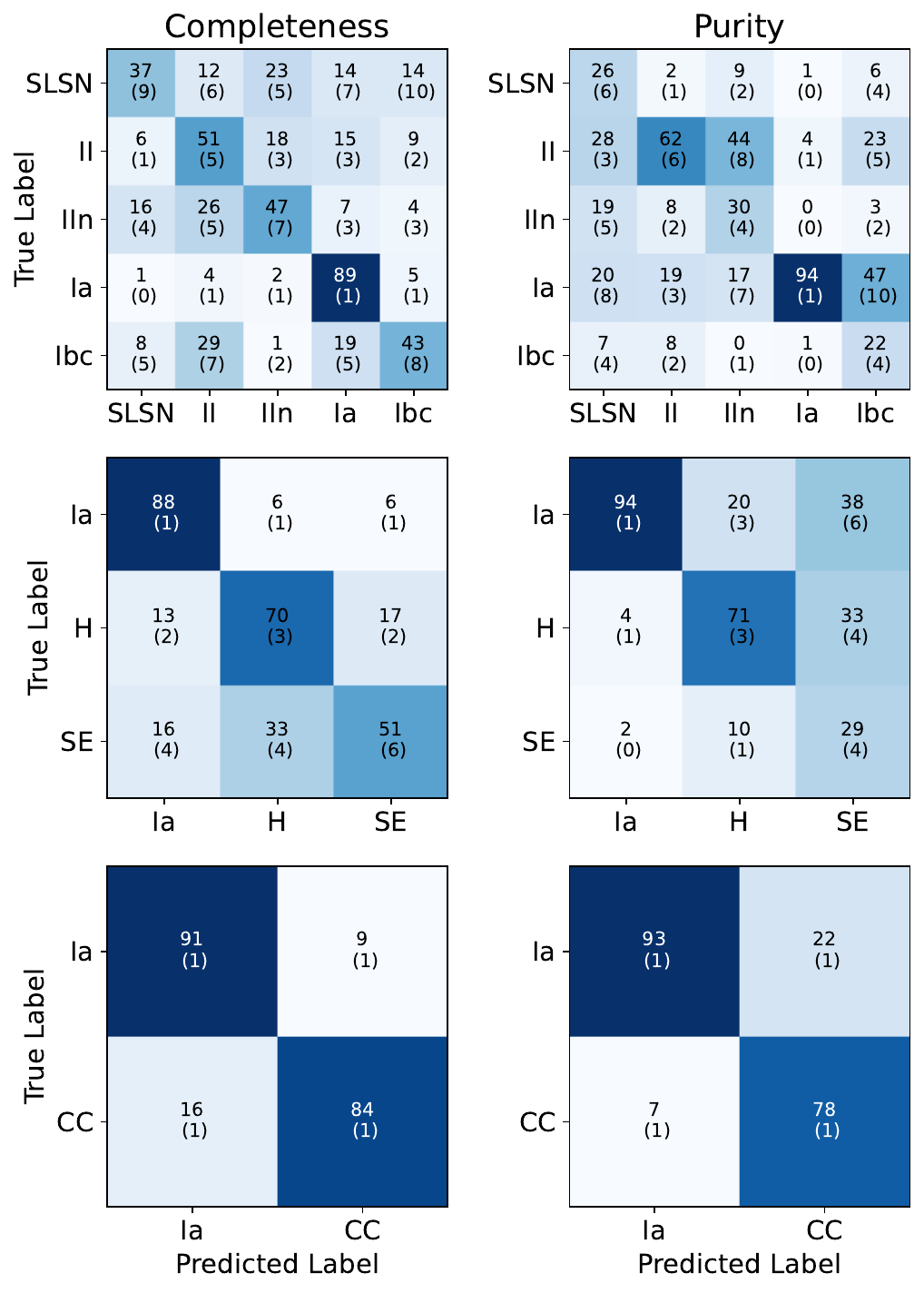}\\
        (b) Light Curve and Galaxy Features, Excluding Redshift
    }
    \caption{Confusion matrices for the classifier using solely light curve features (\textit{left}) and a combination of galaxy and light curve features (\textit{right}), both excluding redshift information. Inclusion of galaxy features does not improve classification for any one class to a statistically significant degree, and the overall performance only moderately increases.}
    \label{fig:conf-gal-2}
\end{figure*}

\begin{figure*}[t]
    \centering
    \parbox{0.45\textwidth}{
        \centering
        \includegraphics[width=\linewidth]{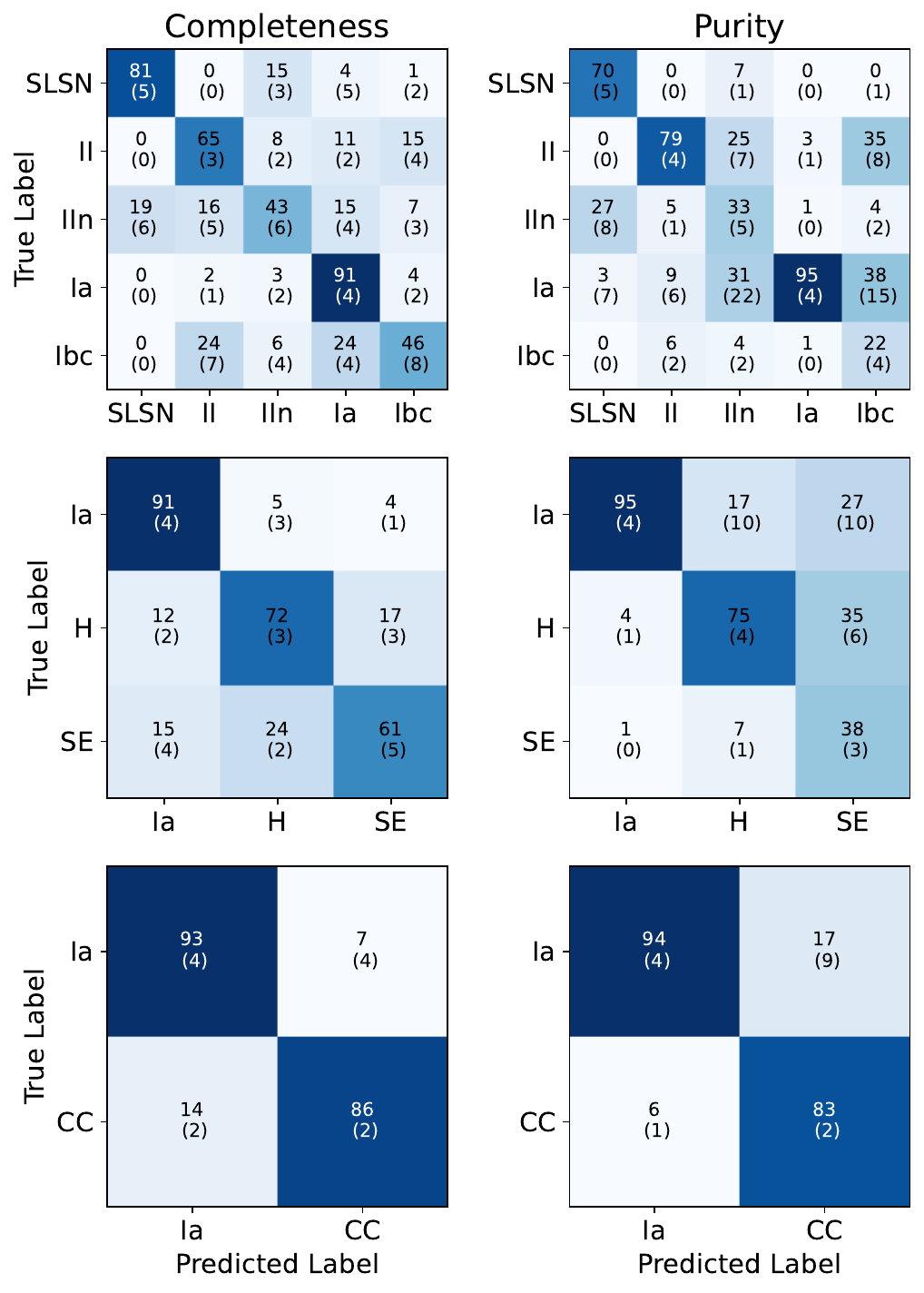}\\
        (a) Light Curve Features, Including Redshift
    }
    \qquad
    \parbox{0.45\textwidth}{
        \centering
        \includegraphics[width=\linewidth]{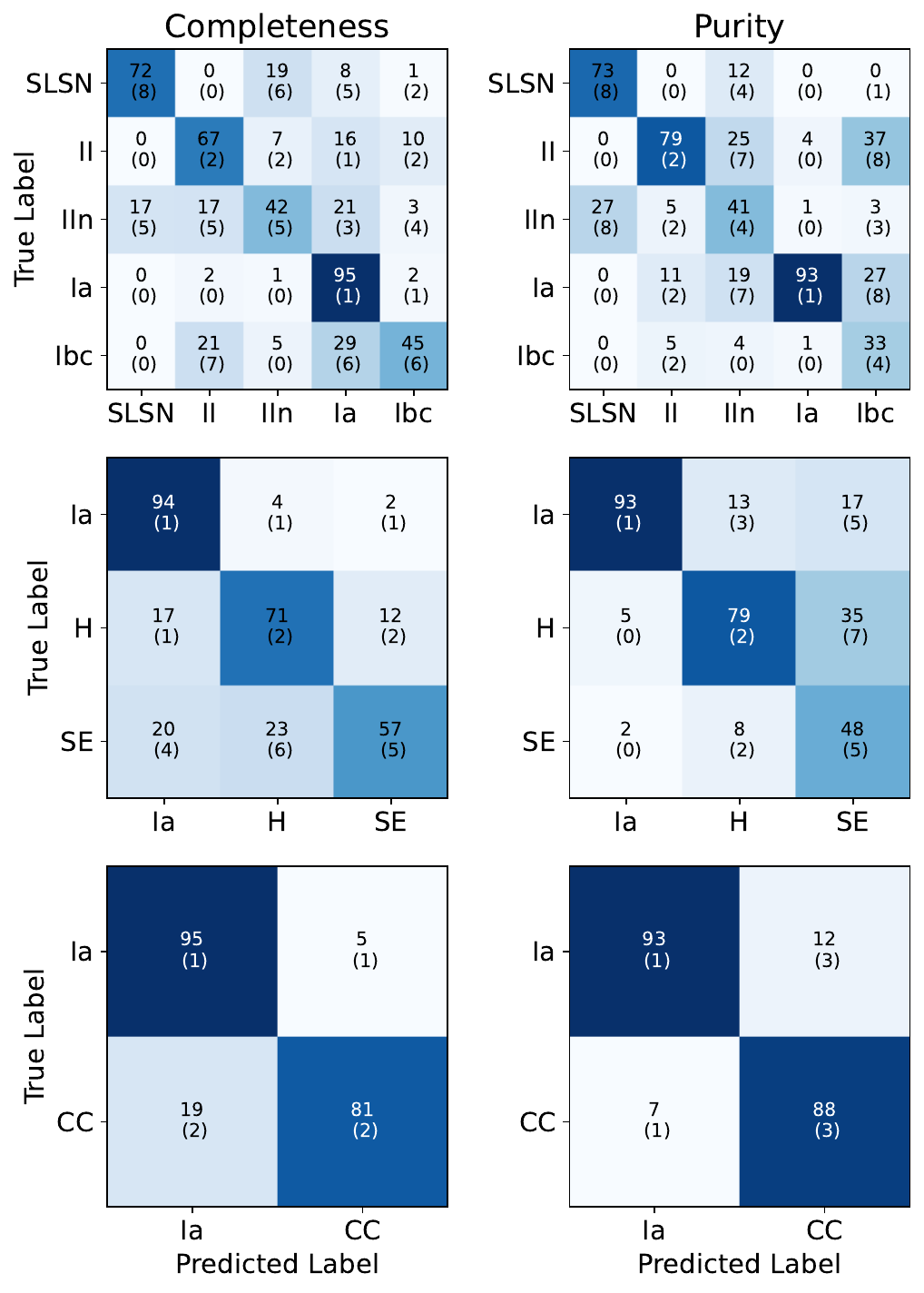}\\
        (b) Light Curve and Galaxy Features, Including Redshift
    }
    \caption{Confusion matrices for the classifier using redshift information, trained using only light curve features (\textit{left}) and with both light curve and galaxy features (\textit{right}). Including host information provides similar overall classification accuracy.}
    \label{fig:conf-gal-3}
\end{figure*}

\subsection{Classification Performance with Host Galaxy Information Only}

We first explore the performance of SN classification from host galaxy information, with and without redshift information (the first two columns of Table~\ref{table:results1}). Our results are primarily visualized in the confusion matrices of Figure~\ref{fig:conf-gal-all}.

In the five-way task without redshift information, we achieve a poor average F1-score of 0.36 across all classes. We find class completeness ranging from $0.28$ (for Type Ibc SNe) to $0.59$ (for Type Ia SNe). We see a much wider spread in purity, from just $0.11$ (for Type Ibc SNe) to $0.86$ (for Type Ia SNe). Unsurprisingly, the majority of incorrectly labelled SNe are dominated by Type Ia SNe, making up $\gtrsim50$\% of each sample. Including redshift information, our average F1-score greatly improves from 0.36 to 0.45. Classification purity and completeness also increase to 0.41 and 0.51, respectively. The class completeness ranges from $0.27$ (for Type IIn SNe) to $0.83$ (for SLSNe). The improvement in the latter is not surprising, as SLSNe are intrinsically luminous and can therefore be uniquely found at higher redshift for a magnitude-limited survey like PS1-MDS (and will similarly be true in LSST). The spread in purity is still quite high, ranging from $0.09$ (for Type Ibc SNe) to $0.91$ (for Type Ia SNe). If we impose a confidence cutoff ($p>0.8$), we find a substantial improvement in overall performance, with the (unweighted) average purity increasing from $0.32$ to $0.41$, and the completeness from $0.40$ to $0.50$, suggesting that purer samples of SNe can be collected with these cuts. Type Ia purity, in particular, increases to $0.94$, suggesting that a highly pure sample of cosmological Type Ia SNe can be selected from host galaxy properties alone. Similarly, SLSNe and Type II SNe can achieve relatively high purity ($>0.7$) with high thresholds ($p>0.8$) when redshift information is included. Unfortunately, the purity of Type IIn and Type Ibc SNe remain low ($\lesssim0.2$) even at this high cutoff.

In the three-way classification task, the galaxy-only classifier achieves a notably higher F1-score of 0.49 even when excluding redshift information. The  average purity (0.46) and completeness (0.52) also increase. Here, again, the minority classes (stripped-envelope and H-rich SNe) suffer from generally low purity scores (0.14 and 0.36, respectively), while Type Ia SNe can be classified with a high purity of 0.88. Including redshift information, the purity and completeness again show positive trends for each class, particularly for stripped-envelope SNe, whose purity increases from 0.14 to 0.22. The F1-score, again significantly improves from $0.49$ to $0.54$.

In the binary classification task (Type Ia vs CCSN), we find average class purity of $0.64$ and completeness of 0.67.  This final classifier is most readily compared to literature studies. The binary average accuracy of our classifier (0.66) is very similar to that of GHOST \citep{gagliano2021ghost}, which reported a class-average accuracy of $0.68$. Including redshift information, we increase this accuracy to $\simeq0.74$, showing state-of-the-art performance in the binary classification task. 

Similarly, in the three-way classification task, we find that our classifier outperforms state-of-the-art in the literature. In particular, we compare the performance of our host-only classifications to those of \cite{gagliano2023first}. \cite{gagliano2023first} uses early (within three days of detection) light curve information, redshift and host information to classify SNe. They find, using the same three-way classification schema, a class-balanced F1-score of $0.48$. Without redshift information, we achieve an F1 score of $0.49$ and with redshift information this substantially increases to $0.54$.

We additionally compare the performance of our classifier on the specialized FLEET classifier for SLSNe. \cite{gomez2020fleet} present three versions of the FLEET classifier: one which only uses host galaxy information and early light curve information (e.g., rise); one which uses host galaxy information and redshift; and one which uses host properties, the full light curve and redshift information. Here, we compare to both the first and second versions. Without redshift information, we achieve low purity in SLSNe (with a maximum purity of $\simeq25$\% with a threshold $p>0.5$). In contrast, FLEET boasts a purity of $\simeq85$\% (with a corresponding completeness of $\simeq20$\%) without redshift information. However, we find that--given redshift information--our classifier notably outperforms that of FLEET. We achieve a purity $\simeq85$\% at a corresponding completeness of $\sim50$\%. FLEET achieves a similar purity at a completeness of $\simeq25$\%.

The key takeaway from the galaxy-based classifier is threefold. First, we are able to successfully isolate a pure ($\simeq0.9$) sample of Type Ia SNe with host galaxy information alone at a reasonably high completeness ($\simeq0.6$) even \textit{without} any redshift information. Second, using redshift information, we can produce reasonable pure samples of Type II SNe and SLSNe, but not Type IIn and Type Ibc SNe. Finally, in the absence of any light curve information, redshift does improve the performance of our classifier for all classes, suggesting that accurate photo-z information will be valuable for rapid classification.

\subsection{Classification Performance Using Light Curve and Galaxy Features}

We next turn our attention to a combined feature set that includes both light curve and host galaxy features. First, we focus on the case where redshift information is not known. Although photometric redshift estimates are often available in current surveys and will be available in the era of LSST, the fraction of catastrophic outlier redshift estimates will be $\simeq0.1$ in the first half of LSST \citep{2018AJ....155....1G}. Transients which are primarily found in intrinsically low-luminosity galaxies (e.g., SLSNe) will have particularly unreliable redshift information. As a baseline comparison, we will contrast our results with both the galaxy-only classifier and a classifier that relies solely on light curve features. Our results can be primarily visualized by the confusion matrices in Figure~\ref{fig:conf-gal-2}.

In the five-way classification task, our ``baseline" classifier with only light curve information achieves an average F1-score of 0.46. This is only marginally improved with the inclusion of galaxy information, increasing to an average F1-score of 0.49. Although the purity and completeness of all classes improve with the addition of galaxy information, although most are not statistically significant ($>1\sigma$). Notably, the purity (0.20 to 0.25), completeness (0.27 to 0.33) and F1-score (0.23 to 0.29) of SLSNe substantially improve, with the F1-score having a statistically significant increase. This is not surprising, as SLSNe are known to prefer unusually low-mass galaxies. The contextual information of their hosts is therefore likely a useful feature. Imposing a higher confidence threshold ($p>0.8$) notably improves the purity of Type II SNe (from 0.62 to 0.70), Type IIn SNe (from 0.29 to 0.52), and Type Ibc SNe (from 0.22 to 0.50) when including galaxy information. These improvements are \textit{not} seen for Type IIn and Type Ibc SNe when excluding galaxy information. 

In both the three-way and binary classification tasks, we see even greater performance improvement. In the three-way task, including galaxy information improves our average F1-score to 0.65, compared to the light curve-only baseline of 0.60; this is a $>3\sigma$ performance improvement. We again see improvement in purity, completeness and F1-scores for all classes. This improvement is substantial for both Type II and stripped-envelope SNe, with the F1-score improving from 0.66 to 0.71 and from 0.30 to 0.38, respectively. For the binary classification, the average F1-score improves from 0.84 to 0.86 ($>1\sigma$), with both Type Ia purity and CC completeness most notably improving (both by $>1\sigma$).

Finally, we explore how galaxy information impacts classification performance when redshift is known. Across all three tasks (five-, three- and binary classifications), there is no statistical difference of average F1-scores between classifier with and without galaxy information (see Fig.~\ref{fig:conf-gal-3}). The only statistically significant difference ($>1\sigma$) is an improvement on the purity of Type Ibc SNe (from 0.24 to 0.32) when including galaxy information, although this comes with a decrease in completeness (from 0.47 to 0.42, within $1\sigma$ uncertainties). We note that including galaxy information leads to higher SN Ia completeness (0.95 vs 0.98; a $>2\sigma$ result).  As expected, a high-confidence cutoff ($p>0.8$), improves accuracy, completeness and purity across all classification tasks with or without galaxy information. To understand if the relative unimportance of host galaxy features is due to the imbalance of host vs light curve features, we train a simple random forest to perform a five-way classification, which allows us to calculate a relative ``importance" score for all features. We find that \textit{no} galaxy features appear in the top 10 important features, and five appear in the upper half of all features. This is in contrast to the importance of features calculated without redshift information. There, five of the top ten important features come from host galaxies. 

In short, contextual host galaxy information improves classification performance when redshift is \textit{not} known. However, when redshift is known, galaxy information does \textit{not} significantly improve classification performance. As a caveat, host galaxy information does improve classification performance for Type Ia SNe (with or without redshift information) when a high confidence threshold is used. These findings hold true for complete light curves, but may change in the case where partial light curve information is known. For example, \cite{gagliano2023first} found that host information led to a $\simeq2\sigma$ increase in binary classification accuracy when utilizing redshift information.

\begin{deluxetable*}{lcccccccc}
\tablecaption{Classification probabilities of all SN-like objects in PS1-MDS. A complete version of this Table is available online. \label{tab:tab3}}
\tablehead{\colhead{Name} & \colhead{$p(\mathrm{Ia})$} & \colhead{$p(\mathrm{CC})$} & \colhead{$p(\mathrm{H\mbox{-}rich})$} & \colhead{$p(\mathrm{H\mbox{-}poor})$} & \colhead{$p(\mathrm{SLSN})$} & \colhead{$p(\mathrm{II})$} & \colhead{$p(\mathrm{IIn})$} & \colhead{$p(\mathrm{Ibc})$}}
\startdata
PSc000012 & $1.00 \pm {<}0.01$ & $0.00 \pm {<}0.01$ & $0.00 \pm {<}0.01$ & $0.00 \pm {<}0.01$ & $0.00 \pm {<}0.01$ & $0.00 \pm {<}0.01$ & $0.00 \pm {<}0.01$ & $0.00 \pm {<}0.01$ \\
PSc000013 & $0.98 \pm 0.02$ & $0.02 \pm 0.02$ & $0.01 \pm 0.01$ & $0.01 \pm 0.01$ & $0.00 \pm 0.00$ & $0.01 \pm 0.01$ & $0.00 \pm {<}0.01$ & $0.01 \pm 0.01$ \\
PSc000015 & $1.00 \pm {<}0.01$ & $0.00 \pm {<}0.01$ & $0.00 \pm {<}0.01$ & $0.00 \pm {<}0.01$ & $0.00 \pm {<}0.01$ & $0.00 \pm {<}0.01$ & $0.00 \pm {<}0.01$ & $0.00 \pm {<}0.01$ \\
$\cdots$ & $\cdots$ & $\cdots$ & $\cdots$ & $\cdots$ & $\cdots$ & $\cdots$ & $\cdots$ & $\cdots$ \\
PSc590246 & $0.00 \pm 0.00$ & $1.00 \pm 0.00$ & $1.00 \pm 0.00$ & $0.00 \pm 0.00$ & $0.00 \pm 0.00$ & $0.67 \pm 0.18$ & $0.33 \pm 0.18$ & $0.00 \pm 0.00$ \\
PSc590248 & $0.99 \pm 0.01$ & $0.01 \pm 0.01$ & $0.01 \pm 0.01$ & $0.00 \pm 0.00$ & $0.00 \pm 0.00$ & $0.01 \pm 0.01$ & $0.00 \pm 0.00$ & $0.00 \pm 0.00$ \\
PSc590260 & $0.85 \pm 0.08$ & $0.15 \pm 0.08$ & $0.14 \pm 0.09$ & $0.00 \pm 0.01$ & $0.00 \pm 0.00$ & $0.13 \pm 0.07$ & $0.01 \pm 0.02$ & $0.00 \pm 0.01$ \\
PSc590263 & $0.00 \pm 0.00$ & $1.00 \pm 0.00$ & $0.51 \pm 0.42$ & $0.49 \pm 0.42$ & $0.44 \pm 0.42$ & $0.08 \pm 0.09$ & $0.43 \pm 0.38$ & $0.04 \pm 0.05$ \\
\enddata
\end{deluxetable*}

\section{Classifications of the PS1-MDS Sample}\label{sec:ps1}

We use our redshift-independent classifier which uses both host and light curve information to classify the full set of 4407 SN-like transients from the PS1-MDS originally presented in \cite{villar2020superraenn} and \cite{hosseinzadeh2020photometric}. This sample includes all objects which are not spectroscopically classified and not identified as variables or otherwise ``bad" objects. The full classifications are provided in Table~\ref{tab:tab3}.

\begin{deluxetable*}{ccccc}
\tabletypesize{\footnotesize}
\tablecolumns{5}
\tablecaption{ Expected and actual agreements between classifiers. \label{table:agree}}
\tablehead{
\colhead{Class} & \colhead{SuperRAENN (expected)} & \colhead{SuperRAENN (actual)} & \colhead{SuperPhot (expected)} & \colhead{SuperPhot (actual)}}
 \startdata
SLSNe & 0.14 & 0.06 & 0.16 & 0.04 \\
SN II & 0.48 & 0.43 & 0.52 & 0.41 \\
SN IIn & 0.15 & 0.08 & 0.13 & 0.09 \\
SN Ia & 0.85 & 0.88 & 0.78 & 0.85 \\
SN Ibc & 0.07 & 0.24 & 0.07 & 0.19 \\
 \enddata
\end{deluxetable*}

\begin{figure*}[t]
\centering
\includegraphics[width=0.8\textwidth]{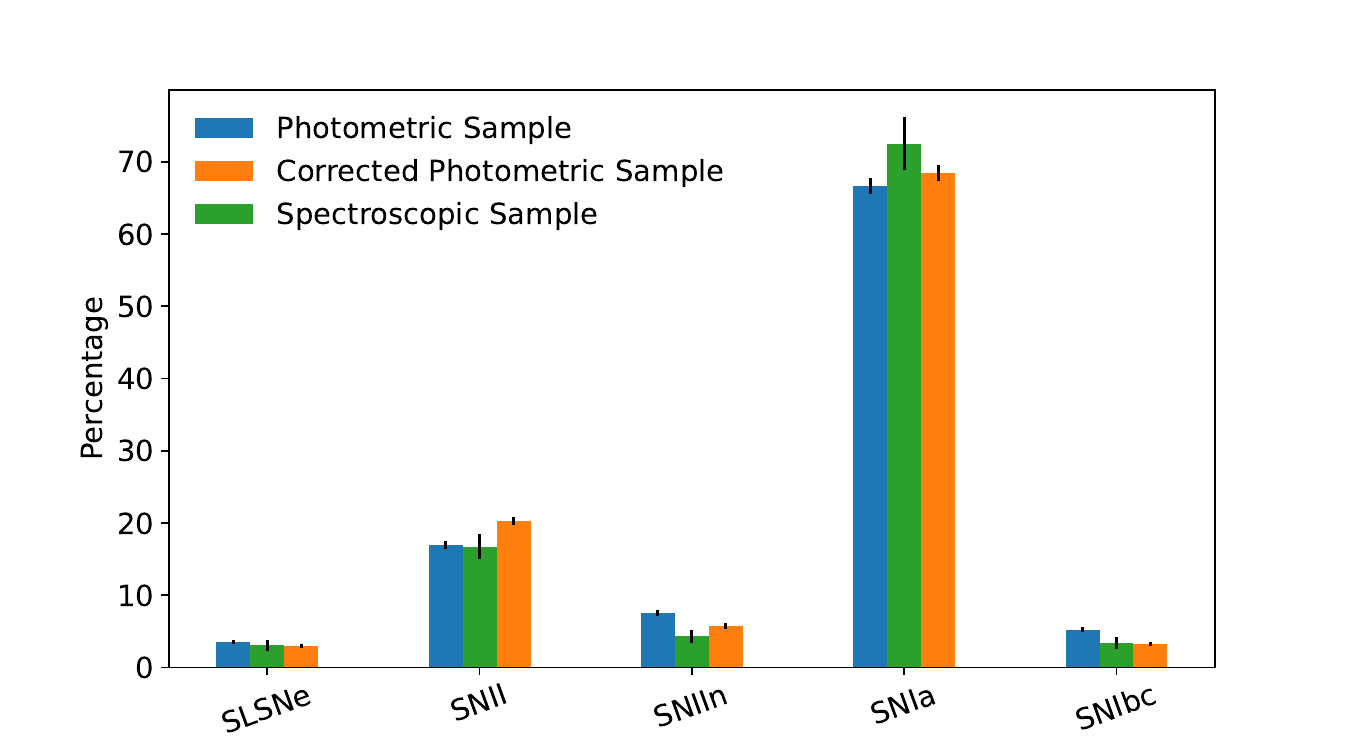}
\caption{Breakdown of SN subclasses in our new photometrically classified set of SNe without redshift information. Compared to the spectroscopic sample (green), our classifier (blue) predicts a smaller fraction of Type Ia and Type II SNe and a larger fraction of SLSNe and Type Ibc SNe. However, using our purity matrix (see text), we can correct our classification breakdowns accounting for expected mis-identifications (orange). Note that errorbars are shown in black and calculated using Poisson uncertainties. We see that, following this correction, our classes breakdowns are largely consistent with the spectroscopic sample. This indicates that although our classifier is biased, it is biased in a way which can be understand via our training set and can likely be carefully corrected for  population-level studies. \label{fig:hist}}
\end{figure*}

The class fractions, as derived by our new classifier, are shown in Figure~\ref{fig:hist}. Compared to the spectroscopic dataset, we find an under representation of the two majority classes (Type Ia and Type II SNe) in our photometric classifications, and an over representation of the minority classes (SLSNe and Type Ibc). \cite{hosseinzadeh2020photometric} showed that the expected number of misclassifications can be used to correct the expected class breakdown (by dotting the purity matrix with the final classifications). When we apply this correction to our final class breakdown, we do recover a five-way class breakdown similar to the spectroscopic sample, lending credence to the idea that we correctly capture the biases of our imperfect classifier, even when applied to a new test set. 

We more quantitatively compare our results to the SNe photometrically classified in \cite{villar2020superraenn} and \cite{hosseinzadeh2020photometric} by analyzing the \textit{agreement} ($A$) between these three classifiers. For a given SN class, the agreement between two classifiers is the number of events which have been identified as the same class by both classifiers divided by the size of the class in the older classifier. \cite{hosseinzadeh2020photometric} showed that the agreement  between two classifiers, assuming independent biases, can be calculated as:

\begin{equation}
    A = P^T C
\end{equation}
where $P$ is the purity matrix of the new classifier and $C$ is the completeness matrix of the old classifiers. Using this, the expected agreement for each class and either classifier is shown in Table~\ref{table:agree}. We will note that the definition of agreement listed above assumes that the two classifiers are independent, which is not true in our case, as both \texttt{SuperRAENN} and our new classifier use the same feature set. This may lead to more agreement than expected between the classifiers. We also note that both \texttt{SuperRAENN} and \texttt{SuperPhot} utilize redshift information, which is not used in our new classifier. 

Averaged across the five SN classes, we find 67\% agreement between the new classifications and those from \texttt{SuperRAENN}, and 59\% agreement between our new classifier and \texttt{SuperPhot}. For each SN class, the agreement between classifiers ranges from $\simeq5-90$\%. Our classifier shows roughly expected agreement for both Type Ia and Type II classes. Type Ia SNe, in particular, have strong agreement between the three classifiers ($80-85$\%), giving high confidence in the purity of our Type Ia sample. Type Ibc SNe have notably higher agreement ($\simeq20$\%) than expected. Interestingly, this is similar to what was found in \cite{hosseinzadeh2020photometric} when comparing \texttt{SuperPhot} to \texttt{SuperRAENN}. Type IIn and SLSNe, on the other hand, have much smaller agreement than expected by a factor of $2-4$; however, these two classes are commonly identified as the other.

We draw further attention to the rare class of SLSNe. Most objects classified by \texttt{SuperRAENN} or \texttt{SuperPhot} as SLSNe were classified as Type Ia SNe by the new classifier. Interestingly, of the $\simeq150$ SLSNe identified by the three classifiers, only one object is identified as a SLSN by all three -- PSc010186. Examination of the light curve and a cross-match with known active galactic nuclei suggest that object is a $z\simeq1$ active galactic nucleus undergoing regular, long-term variation \citep{hsu2022photometrically}. No objects are classified as SNe by our new classifier and \textit{one} of the two original classifiers. We also note two other events presented in \cite{hsu2022photometrically} that are not classified as SLSNe by our new classifier: PSc000036 (classified as a Type Ia) and PSc000553a (classified as a Type IIn). The former is the highest redshift ($z=2.026$) and brightest ($M_\mathrm{g}=-24$) object in the sample of \cite{hsu2022photometrically}, which likely led to the initial classification.  We take these results as an important warning: without redshift information, our photometric classifiers seemingly fail to produce \textit{consistent} and pure samples of SLSNe. Furthermore, accurate AGN classifiers may be particularly important in isolating SLSNe photometrically.

%


\begin{figure}[t]
\includegraphics[width=0.45\textwidth]{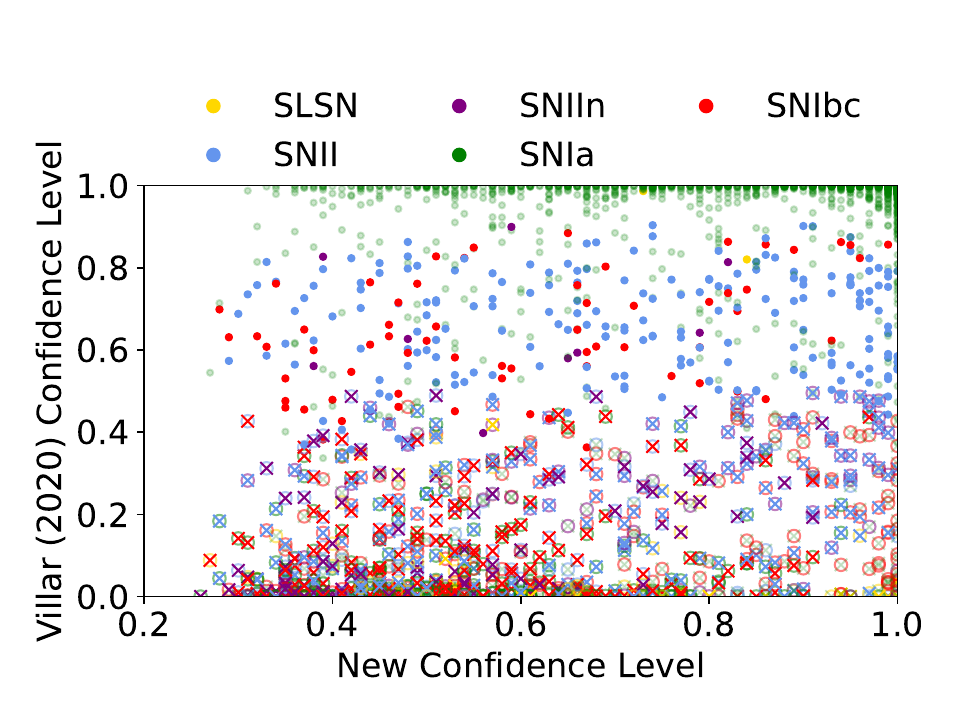}
\caption{Comparison of SN classifications from \cite{villar2020superraenn} and this work. On the abscissa, we plot the confidence level of the \cite{villar2018superluminous} classifier, while the ordinate shows the confidence of the new classifier for the \textit{same class}. Objects plotted as a point have the same classification with both methods; objects plotted as an `x' have different classifications with either method. The color of the central point (`x' or `o') identifies the label from this paper. In the case of mismatches (the `x' objects), the colored circle represents the label from \cite{villar2020superraenn}. \label{fig:misclass}}
\end{figure}

Finally, we visualize our agreement in Figure~\ref{fig:misclass}. In general, Type Ia SNe tend to dominate the sample of objects for which both our new classifier and \cite{villar2020superraenn} confidently agree. It is reassuring that all disagreements occur with the \textit{new} confidence level $<0.5$, and are evenly distributed across the classes. However, a small sample of events have a highly confident classification from \cite{villar2020superraenn} despite disagreement with the new classifier. In particular, 14 events have a confidence score of $<0.05$ in this work and $>0.95$ for the same class. Of these, six were originally classified as SLSNe and are now labelled as Type Ia SNe; one of these is the aforementioned PSc310006. An additional six were originally Type Ia SNe, and are now labelled as CCSNe (Type II and Ibc). 

A key takeaway of this analysis is that expected classification accuracies and ``agreement" between photometric classifiers are highly variable, and any population-level studies of photometrically classified SNe should take care to understand underlying biases and misclassifications in the observed SNe population.


\section{Conclusions}\label{sec:conclusions}

We have presented an analysis on the impact of galaxy properties on the photometric classification of SNe, focusing on CCSN subtypes. We present our key findings and products below:

\begingroup
\begin{enumerate}[topsep=0pt, itemsep=10pt, parsep=1pt, after=\vspace{0pt}]
    \item In corroboration with previous results, we find that the observable properties of host galaxies visibly correlate with SN type. We are able to successfully distinguish CC SNe from Type Ia SNe with $\simeq70$\% accuracy, matching similar studies in the literature.
    \item We are able to produce relatively pure ($>90$\%) samples of Type Ia SNe using host galaxy information alone, with or without redshift information. We are able to produce reasonable pure samples ($>70$\%) samples of SLSNe and Type II SNe when using host galaxy and redshift information.
    \item We present the first application of the weighted hierarchical cross-entropy score (see \citealt{villar2023hierarchical}), which better accounts for the hierarchical taxonomy of SN classification. 
    \item Given redshift, we find that contextual information -- i.e., the host galaxy observable properties -- does not necessarily aid in SN classification when a full light curve is available. Host galaxy information is most helpful in improving the classification accuracy of Type Ia and Type II SNe, but does not necessarily increase the accuracy of classification for other subtypes.
    \item Given no redshift information, host galaxy features consistently improve the classification accuracy of SNe across all subtypes. Galaxy information can greatly increase the completeness of Type Ia samples.
    \item Finally, we present an updated classification set for the PS1-MDS set of SN-like transients without redshift information. We explore the agreement between these classifications and those originally presented in \citealt{hosseinzadeh2020photometric} and \citealt{villar2020superraenn}, finding strong agreement in some classes (Type Ia, Type Ibc) and weaker than expected agreement in others (most notably, SLSNe).
\end{enumerate}
\endgroup

In the era of LSST, thousands of new SNe are expected to be discovered nightly. Our results show that host galaxy information alone can distinguish between core-collapse and thermonuclear SNe, even in the absence of redshift information. However, accurate photometric redshift estimates from host galaxies will likely greatly increase our ability to rapidly classify further subtypes (particularly Type Ia and SLSNe). Given the expected on-sky density of galaxies ($\sim10^4$ per square degree), reliable host galaxy association techniques, such as \texttt{GHOST}, are essential in utilizing this contextual information.

Finally, the variation on agreement between classifiers explored here emphasizes the need for methodologies to calibrate  classification confidence and build reliable populations of photometrically-classified SNe. SLSNe stand out as particularly challenging. Here, we find just a single object labelled as a SLSN by all three classifiers out of dozens of events. Future work should investigate the use of probabilistic labels into population-level inference, e.g., similar to \cite{kunz2007bayesian}.

\begin{acknowledgements}
VAV acknowledges support by the NSF through grant AST-2108676. This work is also supported by the NSF through PHY-2019786 (the NSF AI Institute for Artificial Intelligence and Fundamental Interactions). Some of the computations in this paper were run on the Cannon cluster supported by the FAS Division of Science, Research Computing Group at Harvard University. The Pan-STARRS1 Surveys (PS1) and the PS1 public science archive have been made possible through contributions by the Institute for Astronomy, the University of Hawaii, the Pan-STARRS Project Office, the Max-Planck Society and its participating institutes, the Max Planck Institute for Astronomy, Heidelberg, and the Max Planck Institute for Extraterrestrial Physics, Garching, Johns Hopkins University, Durham University, the University of Edinburgh, the Queen’s University Belfast, the Harvard-Smithsonian Center for Astrophysics, the Las Cumbres Observatory Global Telescope Network Incorporated, the National Central University of Taiwan, the Space Telescope Science Institute, the National Aeronautics and Space Administration under grant No. NNX08AR22G issued through the Planetary Science Division of the NASA Science Mission Directorate, the National Science Foundation grant No.
AST-1238877, the University of Maryland, Eotvos Lorand University (ELTE), the Los Alamos National Laboratory, and the Gordon and Betty Moore Foundation.
\end{acknowledgements}

\software{
    Astropy \citep{robitaille2013astropy, 2018AJ....156..123A, 2022ApJ...935..167A},
    FLEET \citep{gomez2020fleet},
    Matplotlib \citep{matplotlib},
    NumPy \citep{numpy},
    Superphot \citep{hosseinzadeh2020photometric},
    SuperRAENN \citep{villar2020superraenn},
    Pytorch \citep{pytorch}
}

\bibliography{sample631}{}

@inproceedings{bertinetto2020making,
  title={Making better mistakes: Leveraging class hierarchies with deep networks},
  author={Bertinetto, Luca and Mueller, Romain and Tertikas, Konstantinos and Samangooei, Sina and Lord, Nicholas A},
  booktitle={Proceedings of the IEEE/CVF Conference on Computer Vision and Pattern Recognition},
  pages={12506--12515},
  year={2020}
}

@article{foley2013classifying,
  title={Classifying supernovae using only galaxy data},
  author={Foley, Ryan J and Mandel, Kaisey},
  journal={The Astrophysical Journal},
  volume={778},
  number={2},
  pages={167},
  year={2013},
  publisher={IOP Publishing}
}

@article{kessler2019models,
  title={Models and simulations for the photometric LSST astronomical time series classification challenge (PLAsTiCC)},
  author={Kessler, R and Narayan, G and Avelino, A and Bachelet, E and Biswas, Rahul and Brown, PJ and Chernoff, DF and Connolly, AJ and Dai, M and Daniel, S and others},
  journal={Publications of the Astronomical Society of the Pacific},
  volume={131},
  number={1003},
  pages={094501},
  year={2019},
  publisher={IOP Publishing}
}

@article{gagliano2021ghost,
  title={GHOST: Using Only Host Galaxy Information to Accurately Associate and Distinguish Supernovae},
  author={Gagliano, Alex and Narayan, Gautham and Engel, Andrew and Kind, Matias Carrasco and LSST Dark Energy Science Collaboration and others},
  journal={The Astrophysical Journal},
  volume={908},
  number={2},
  pages={170},
  year={2021},
  publisher={IOP Publishing}
}

@article{baldeschi2020star,
  title={Star Formation and Morphological Properties of Galaxies in the Pan-STARRS 3$\pi$ Survey. I. A Machine-learning Approach to Galaxy and Supernova Classification},
  author={Baldeschi, A and Miller, A and Stroh, M and Margutti, R and Coppejans, DL},
  journal={The Astrophysical Journal},
  volume={902},
  number={1},
  pages={60},
  year={2020},
  publisher={IOP Publishing}
}

@article{berger2010short,
  title={A short gamma-ray burst “No-host” problem? Investigating large progenitor offsets for short GRBs with optical afterglows},
  author={Berger, E},
  journal={The Astrophysical Journal},
  volume={722},
  number={2},
  pages={1946},
  year={2010},
  publisher={IOP Publishing}
}

@article{bloom2002observed,
  title={The observed offset distribution of gamma-ray bursts from their host galaxies: a robust clue to the nature of the progenitors},
  author={Bloom, Joshua S and Kulkarni, Shrinivas R and Djorgovski, S George},
  journal={The Astronomical Journal},
  volume={123},
  number={3},
  pages={1111},
  year={2002},
  publisher={IOP Publishing}
}

@article{hudelot2012vizier,
  title={VizieR online data catalog: The CFHTLS survey (T0007 release)(Hudelot+ 2012)},
  author={Hudelot, P and Cuillandre, J-Ch and Withington, K and Goranova, Y and McCracken, H and Magnard, F and Mellier, Y and Regnault, N and Betoule, M and Aussel, H and others},
  journal={VizieR Online Data Catalog},
  pages={II--317},
  year={2012}
}

@article{gomez2020fleet,
  title={FLEET: A Redshift-agnostic Machine Learning Pipeline to Rapidly Identify Hydrogen-poor Superluminous Supernovae},
  author={Gomez, Sebastian and Berger, Edo and Blanchard, Peter K and Hosseinzadeh, Griffin and Nicholl, Matt and Villar, V Ashley and Yin, Yao},
  journal={The Astrophysical Journal},
  volume={904},
  number={1},
  pages={74},
  year={2020},
  publisher={IOP Publishing}
}

@article{villar2020superraenn,
  title={SuperRAENN: A Semisupervised Supernova Photometric Classification Pipeline Trained on Pan-STARRS1 Medium-Deep Survey Supernovae},
  author={Villar, V Ashley and Hosseinzadeh, Griffin and Berger, Edo and Ntampaka, Michelle and Jones, David O and Challis, Peter and Chornock, Ryan and Drout, Maria R and Foley, Ryan J and Kirshner, Robert P and others},
  journal={The Astrophysical Journal},
  volume={905},
  number={2},
  pages={94},
  year={2020},
  publisher={IOP Publishing}
}

@article{muthukrishna2019rapid,
  title={RAPID: early classification of explosive transients using deep learning},
  author={Muthukrishna, Daniel and Narayan, Gautham and Mandel, Kaisey S and Biswas, Rahul and Hlo{\v{z}}ek, Ren{\'e}e},
  journal={Publications of the Astronomical Society of the Pacific},
  volume={131},
  number={1005},
  pages={118002},
  year={2019},
  publisher={IOP Publishing}
}

@article{moller2020supernnova,
  title={SuperNNova: an open-source framework for Bayesian, neural network-based supernova classification},
  author={M{\"o}ller, Anais and de Boissi{\`e}re, Thibault},
  journal={Monthly Notices of the Royal Astronomical Society},
  volume={491},
  number={3},
  pages={4277--4293},
  year={2020},
  publisher={Oxford University Press}
}

@article{carrasco2021alert,
  title={Alert classification for the ALeRCE broker system: The real-time stamp classifier},
  author={Carrasco-Davis, Rodrigo and Reyes, Esteban and Valenzuela, Camilo and F{\"o}rster, Francisco and Est{\'e}vez, Pablo A and Pignata, Giuliano and Bauer, Franz E and Reyes, Ignacio and S{\'a}nchez-S{\'a}ez, Paula and Cabrera-Vives, Guillermo and others},
  journal={The Astronomical Journal},
  volume={162},
  number={6},
  pages={231},
  year={2021},
  publisher={IOP Publishing}
}

@article{hosseinzadeh2020photometric,
  title={Photometric Classification of 2315 Pan-STARRS1 Supernovae with Superphot},
  author={Hosseinzadeh, Griffin and Dauphin, Frederick and Villar, V Ashley and Berger, Edo and Jones, David O and Challis, Peter and Chornock, Ryan and Drout, Maria R and Foley, Ryan J and Kirshner, Robert P and others},
  journal={The Astrophysical Journal},
  volume={905},
  number={2},
  pages={93},
  year={2020},
  publisher={IOP Publishing}
}

@article{qu2022photometric,
  title={Photometric Classification of Early-time Supernova Light Curves with SCONE},
  author={Qu, Helen and Sako, Masao},
  journal={The Astronomical Journal},
  volume={163},
  number={2},
  pages={57},
  year={2022},
  publisher={IOP Publishing}
}

@article{boone2021parsnip,
  title={ParSNIP: Generative Models of Transient Light Curves with Physics-enabled Deep Learning},
  author={Boone, Kyle},
  journal={The Astronomical Journal},
  volume={162},
  number={6},
  pages={275},
  year={2021},
  publisher={IOP Publishing}
}

@article{khazov2016flash,
  title={Flash spectroscopy: emission lines from the ionized circumstellar material around< 10-day-old type II supernovae},
  author={Khazov, Daniel and Yaron, Ofer and Gal-Yam, Avishay and Manulis, Ilan and Rubin, Alon and Kulkarni, SR and Arcavi, Iair and Kasliwal, MM and Ofek, EO and Cao, Y and others},
  journal={The Astrophysical Journal},
  volume={818},
  number={1},
  pages={3},
  year={2016},
  publisher={IOP Publishing}
}

@article{childress2013host,
  title={Host Galaxies of Type Ia Supernovae from the Nearby Supernova Factory},
  author={Childress, Michael and Aldering, G and Antilogus, P and Aragon, C and Bailey, S and Baltay, C and Bongard, S and Buton, C and Canto, A and Cellier-Holzem, F and others},
  journal={The Astrophysical Journal},
  volume={770},
  number={2},
  pages={107},
  year={2013},
  publisher={IOP Publishing}
}

@article{schulze2021palomar,
  title={The Palomar Transient Factory core-collapse supernova host-galaxy sample. I. Host-galaxy distribution functions and environment dependence of core-collapse supernovae},
  author={Schulze, Steve and Yaron, Ofer and Sollerman, Jesper and Leloudas, Giorgos and Gal, Amit and Wright, Angus H and Lunnan, Ragnhild and Gal-Yam, Avishay and Ofek, Eran O and Perley, Daniel A and others},
  journal={The Astrophysical Journal Supplement Series},
  volume={255},
  number={2},
  pages={29},
  year={2021},
  publisher={IOP Publishing}
}

@article{habergham2012central,
  title={A central excess of stripped-envelope supernovae within disturbed galaxies},
  author={Habergham, SM and James, PA and Anderson, JP},
  journal={Monthly Notices of the Royal Astronomical Society},
  volume={424},
  number={4},
  pages={2841--2853},
  year={2012},
  publisher={The Royal Astronomical Society}
}

@article{kelly2012core,
  title={Core-collapse supernovae and host galaxy stellar populations},
  author={Kelly, Patrick L and Kirshner, Robert P},
  journal={The Astrophysical Journal},
  volume={759},
  number={2},
  pages={107},
  year={2012},
  publisher={IOP Publishing}
}

@article{ransome2022h,
  title={An H $\alpha$ survey of the host environments of 77 type IIn supernovae within z< 0.02},
  author={Ransome, Conor L and Habergham-Mawson, SM and Darnley, Matt J and James, Phil A and Percival, Sue M},
  journal={Monthly Notices of the Royal Astronomical Society},
  volume={513},
  number={3},
  pages={3564--3576},
  year={2022},
  publisher={Oxford University Press}
}

@article{anderson2012progenitor,
  title={Progenitor mass constraints for core-collapse supernovae from correlations with host galaxy star formation},
  author={Anderson, Joseph P and Habergham, SM and James, PA and Hamuy, M},
  journal={Monthly Notices of the Royal Astronomical Society},
  volume={424},
  number={2},
  pages={1372--1391},
  year={2012},
  publisher={Blackwell Science Ltd Oxford, UK}
}

@article{dong2022physical,
  title={Physical Properties of the Host Galaxies of Ca-rich Transients},
  author={Dong, Yuxin and Milisavljevic, Dan and Leja, Joel and Sarbadhicary, Sumit K and Nugent, Anya E and Margutti, Raffaella and Jacobson-Gal{\'a}n, Wynn V and Polin, Abigail and Banovetz, John and Reynolds, Jack M and others},
  journal={The Astrophysical Journal},
  volume={927},
  number={2},
  pages={199},
  year={2022},
  publisher={IOP Publishing}
}

@article{hsu2022photometrically,
  title={Photometrically-Classified Superluminous Supernovae from the Pan-STARRS1 Medium Deep Survey: A Case Study for Science with Machine Learning-Based Classification},
  author={Hsu, Brian and Hosseinzadeh, Griffin and Villar, V Ashley and Berger, Edo},
  journal={arXiv preprint arXiv:2204.09809},
  year={2022}
}

@article{villar2018superluminous,
  title={Superluminous Supernovae in LSST: Rates, Detection Metrics, and Light-curve Modeling},
  author={Villar, V Ashley and Nicholl, Matt and Berger, Edo},
  journal={The Astrophysical Journal},
  volume={869},
  number={2},
  pages={166},
  year={2018},
  publisher={IOP Publishing}
}

@article{villar2019supernova,
  title={Supernova Photometric Classification Pipelines Trained on Spectroscopically Classified Supernovae from the Pan-STARRS1 Medium-deep Survey},
  author={Villar, VA and Berger, E and Miller, G and Chornock, R and Rest, A and Jones, DO and Drout, MR and Foley, RJ and Kirshner, R and Lunnan, Ragnhild and others},
  journal={The Astrophysical Journal},
  volume={884},
  number={1},
  pages={83},
  year={2019},
  publisher={IOP Publishing}
}

@article{gomez2023first,
  title={The first two years of FLEET: an active search for superluminous supernovae},
  author={Gomez, Sebastian and Berger, Edo and Blanchard, Peter K and Hosseinzadeh, Griffin and Nicholl, Matt and Hiramatsu, Daichi and Villar, V Ashley and Yin, Yao},
  journal={The Astrophysical Journal},
  volume={949},
  number={2},
  pages={114},
  year={2023},
  publisher={IOP Publishing}
}

@article{kingma2014adam,
  title={Adam: A method for stochastic optimization},
  author={Kingma, Diederik P and Ba, Jimmy},
  journal={arXiv preprint arXiv:1412.6980},
  year={2014}
}

@article{gal2016observational,
       author = {{Gal-Yam}, Avishay},
        title = "{Observational and Physical Classification of Supernovae}",
    booktitle = {Handbook of Supernovae},
         year = 2017,
       editor = {{Alsabti}, Athem W. and {Murdin}, Paul},
        pages = {195},
          doi = {10.1007/978-3-319-21846-5_35},
       adsurl = {https://ui.adsabs.harvard.edu/abs/2017hsn..book..195G},
         journal={arXiv preprint arXiv:1611.09353}
}

@article{gagliano2023first,
  title={First Impressions: Early-Time Classification of Supernovae using Host Galaxy Information and Shallow Learning},
  author={Gagliano, Alexander and Contardo, Gabriella and Mackey, Daniel Foreman and Malz, Alex I and Aleo, Patrick D},
  journal={arXiv preprint arXiv:2305.08894},
  year={2023}
}

@article{collaboration2020planck,
  title={Planck 2018 results. VI. Cosmological parameters},
  author={Aghanim, N and others},
  journal={Astron. Astrophys},
  volume={641},
  pages={A6},
  year={2020}
}

@article{villar2023hierarchical,
  title={Hierarchical Cross-entropy Loss for Classification of Astrophysical Transients},
  author={Villar, V Ashley and de Soto, Kaylee and Gagliano, Alex},
  journal={arXiv preprint arXiv:2312.02266},
  year={2023}
}

@ARTICLE{2018AJ....155....1G,
       author = {{Graham}, Melissa L. and {Connolly}, Andrew J. and {Ivezi{\'c}}, {\v{Z}}eljko and {Schmidt}, Samuel J. and {Jones}, R. Lynne and {Juri{\'c}}, Mario and {Daniel}, Scott F. and {Yoachim}, Peter},
        title = "{Photometric Redshifts with the LSST: Evaluating Survey Observing Strategies}",
      journal = {\aj},
         year = 2018,
        month = jan,
       volume = {155},
       number = {1},
          eid = {1},
        pages = {1},
          doi = {10.3847/1538-3881/aa99d4},
archivePrefix = {arXiv},
       eprint = {1706.09507},
 primaryClass = {astro-ph.CO},
       adsurl = {https://ui.adsabs.harvard.edu/abs/2018AJ....155....1G}
}

@article{bruch2021large,
  title={A large fraction of hydrogen-rich supernova progenitors experience elevated mass loss shortly prior to explosion},
  author={Bruch, Rachel J and Gal-Yam, Avishay and Schulze, Steve and Yaron, Ofer and Yang, Yi and Soumagnac, Maayane and Rigault, Mickael and Strotjohann, Nora L and Ofek, Eran and Sollerman, Jesper and others},
  journal={The Astrophysical Journal},
  volume={912},
  number={1},
  pages={46},
  year={2021},
  publisher={IOP Publishing}
}

@article{hakobyan2012supernovae,
  title={Supernovae and their host galaxies-I. The SDSS DR8 database and statistics},
  author={Hakobyan, AA and Adibekyan, V Zh and Aramyan, LS and Petrosian, AR and Gomes, JM and Mamon, GA and Kunth, D and Turatto, M},
  journal={Astronomy \& Astrophysics},
  volume={544},
  pages={A81},
  year={2012},
  publisher={EDP Sciences}
}

@article{leaman2011nearby,
  title={Nearby supernova rates from the Lick Observatory Supernova Search--I. The methods and data base},
  author={Leaman, Jesse and Li, Weidong and Chornock, Ryan and Filippenko, Alexei V},
  journal={Monthly Notices of the Royal Astronomical Society},
  volume={412},
  number={3},
  pages={1419--1440},
  year={2011},
  publisher={Blackwell Publishing Ltd Oxford, UK}
}

@article{kunz2007bayesian,
  title={Bayesian estimation applied to multiple species},
  author={Kunz, Martin and Bassett, Bruce A and Hlozek, Ren{\'e}e A},
  journal={Physical Review D—Particles, Fields, Gravitation, and Cosmology},
  volume={75},
  number={10},
  pages={103508},
  year={2007},
  publisher={APS}
}

@ARTICLE{2022arXiv220902784K,
       author = {{Kisley}, Marina and {Qin}, Yu-Jing and {Zabludoff}, Ann and {Barnard}, Kobus and {Ko}, Chia-Lin},
        title = "{Classifying Transients Using Host Galaxy Photometry}",
      journal = {arXiv e-prints},
         year = 2022,
        month = sep,
          eid = {arXiv:2209.02784},
        pages = {arXiv:2209.02784},
archivePrefix = {arXiv},
       eprint = {2209.02784},
 primaryClass = {astro-ph.GA},
       adsurl = {https://ui.adsabs.harvard.edu/abs/2022arXiv220902784K}
}

@article{qin2022linking,
  title={Linking Extragalactic Transients and Their Host Galaxy Properties: Transient Sample, Multiwavelength Host Identification, and Database Construction},
  author={Qin, Yu-Jing and Zabludoff, Ann and Kisley, Marina and Liu, Yuantian and Arcavi, Iair and Barnard, Kobus and Behroozi, Peter and French, K Decker and McCully, Curtis and Merchant, Nirav},
  journal={The Astrophysical Journal Supplement Series},
  volume={259},
  number={1},
  pages={13},
  year={2022},
  publisher={IOP Publishing}
}

@article{sanchez2021alert,
  title={Alert classification for the ALeRCE broker system: the light curve classifier},
  author={S{\'a}nchez-S{\'a}ez, P and Reyes, I and Valenzuela, C and F{\"o}rster, F and Eyheramendy, S and Elorrieta, F and Bauer, FE and Cabrera-Vives, G and Est{\'e}vez, PA and Catelan, M and others},
  journal={The Astronomical Journal},
  volume={161},
  number={3},
  pages={141},
  year={2021},
  publisher={IOP Publishing}
}

@article{robitaille2013astropy,
  title={Astropy: A community Python package for astronomy},
  author={Robitaille, Thomas P and Tollerud, Erik J and Greenfield, Perry and Droettboom, Michael and Bray, Erik and Aldcroft, Tom and Davis, Matt and Ginsburg, Adam and Price-Whelan, Adrian M and Kerzendorf, Wolfgang E and others},
  journal={Astronomy \& Astrophysics},
  volume={558},
  pages={A33},
  year={2013},
  publisher={EDP Sciences}
}

@Article{matplotlib,
  Author    = {Hunter, J. D.},
  Title     = {Matplotlib: A 2D graphics environment},
  Journal   = {Computing in Science \& Engineering},
  Volume    = {9},
  Number    = {3},
  Pages     = {90--95},
  publisher = {IEEE COMPUTER SOC},
  doi       = {10.1109/MCSE.2007.55},
  year      = 2007
}

@Article{numpy,
 title         = {Array programming with {NumPy}},
 author        = {Charles R. Harris and K. Jarrod Millman and St{\'{e}}fan J.
                 van der Walt and Ralf Gommers and Pauli Virtanen and David
                 Cournapeau and Eric Wieser and Julian Taylor and Sebastian
                 Berg and Nathaniel J. Smith and Robert Kern and Matti Picus
                 and Stephan Hoyer and Marten H. van Kerkwijk and Matthew
                 Brett and Allan Haldane and Jaime Fern{\'{a}}ndez del
                 R{\'{i}}o and Mark Wiebe and Pearu Peterson and Pierre
                 G{\'{e}}rard-Marchant and Kevin Sheppard and Tyler Reddy and
                 Warren Weckesser and Hameer Abbasi and Christoph Gohlke and
                 Travis E. Oliphant},
 year          = {2020},
 month         = sep,
 journal       = {Nature},
 volume        = {585},
 number        = {7825},
 pages         = {357--362},
 doi           = {10.1038/s41586-020-2649-2},
 publisher     = {Springer Science and Business Media {LLC}},
 url           = {https://doi.org/10.1038/s41586-020-2649-2}
}

@ARTICLE{pytorch,
       author = {{Paszke}, Adam and {Gross}, Sam and {Massa}, Francisco and {Lerer}, Adam and {Bradbury}, James and {Chanan}, Gregory and {Killeen}, Trevor and {Lin}, Zeming and {Gimelshein}, Natalia and {Antiga}, Luca and {Desmaison}, Alban and {K{\"o}pf}, Andreas and {Yang}, Edward and {DeVito}, Zach and {Raison}, Martin and {Tejani}, Alykhan and {Chilamkurthy}, Sasank and {Steiner}, Benoit and {Fang}, Lu and {Bai}, Junjie and {Chintala}, Soumith},
        title = "{PyTorch: An Imperative Style, High-Performance Deep Learning Library}",
      journal = {arXiv e-prints},
         year = 2019,
        month = dec,
          eid = {arXiv:1912.01703},
        pages = {arXiv:1912.01703},
          doi = {10.48550/arXiv.1912.01703},
archivePrefix = {arXiv},
       eprint = {1912.01703},
 primaryClass = {cs.LG},
       adsurl = {https://ui.adsabs.harvard.edu/abs/2019arXiv191201703P}
}

@article{swann20194most,
  title={4MOST Consortium Survey 10: The time-domain extragalactic survey (TiDES)},
  author={Swann, Elizabeth and Sullivan, Mark and Carrick, Jonathan and Hoenig, Sebastian and Hook, Isobel and Kotak, Rubina and Maguire, Kate and McMahon, Richard and Nichol, Robert and Smartt, Stephen},
  journal={arXiv preprint arXiv:1903.02476},
  year={2019}
}

@article{kasliwal2012calcium,
  title={Calcium-rich gap transients in the remote outskirts of galaxies},
  author={Kasliwal, Mansi M and Kulkarni, SR and Gal-Yam, Avishay and Nugent, Peter E and Sullivan, Mark and Bildsten, Lars and Yaron, Ofer and Perets, Hagai B and Arcavi, Iair and Ben-Ami, Sagi and others},
  journal={The Astrophysical Journal},
  volume={755},
  number={2},
  pages={161},
  year={2012},
  publisher={IOP Publishing}
}

@article{jacobson2024final,
  title={Final Moments II: Observational Properties and Physical Modeling of CSM-Interacting Type II Supernovae},
  author={Jacobson-Gal{\'a}n, WV and Dessart, L and Davis, KW and Kilpatrick, CD and Margutti, R and Foley, RJ and Chornock, R and Terreran, G and Hiramatsu, D and Newsome, M and others},
  journal={arXiv preprint arXiv:2403.02382},
  year={2024}
}

@ARTICLE{2018AJ....156..123A,
       author = {{Astropy Collaboration} and {Price-Whelan}, A.~M. and {Sip{\H{o}}cz}, B.~M. and {G{\"u}nther}, H.~M. and {Lim}, P.~L. and {Crawford}, S.~M. and {Conseil}, S. and {Shupe}, D.~L. and {Craig}, M.~W. and {Dencheva}, N. and {Ginsburg}, A. and {VanderPlas}, J.~T. and {Bradley}, L.~D. and {P{\'e}rez-Su{\'a}rez}, D. and {de Val-Borro}, M. and {Aldcroft}, T.~L. and {Cruz}, K.~L. and {Robitaille}, T.~P. and {Tollerud}, E.~J. and {Ardelean}, C. and {Babej}, T. and {Bach}, Y.~P. and {Bachetti}, M. and {Bakanov}, A.~V. and {Bamford}, S.~P. and {Barentsen}, G. and {Barmby}, P. and {Baumbach}, A. and {Berry}, K.~L. and {Biscani}, F. and {Boquien}, M. and {Bostroem}, K.~A. and {Bouma}, L.~G. and {Brammer}, G.~B. and {Bray}, E.~M. and {Breytenbach}, H. and {Buddelmeijer}, H. and {Burke}, D.~J. and {Calderone}, G. and {Cano Rodr{\'\i}guez}, J.~L. and {Cara}, M. and {Cardoso}, J.~V.~M. and {Cheedella}, S. and {Copin}, Y. and {Corrales}, L. and {Crichton}, D. and {D'Avella}, D. and {Deil}, C. and {Depagne}, {\'E}. and {Dietrich}, J.~P. and {Donath}, A. and {Droettboom}, M. and {Earl}, N. and {Erben}, T. and {Fabbro}, S. and {Ferreira}, L.~A. and {Finethy}, T. and {Fox}, R.~T. and {Garrison}, L.~H. and {Gibbons}, S.~L.~J. and {Goldstein}, D.~A. and {Gommers}, R. and {Greco}, J.~P. and {Greenfield}, P. and {Groener}, A.~M. and {Grollier}, F. and {Hagen}, A. and {Hirst}, P. and {Homeier}, D. and {Horton}, A.~J. and {Hosseinzadeh}, G. and {Hu}, L. and {Hunkeler}, J.~S. and {Ivezi{\'c}}, {\v{Z}}. and {Jain}, A. and {Jenness}, T. and {Kanarek}, G. and {Kendrew}, S. and {Kern}, N.~S. and {Kerzendorf}, W.~E. and {Khvalko}, A. and {King}, J. and {Kirkby}, D. and {Kulkarni}, A.~M. and {Kumar}, A. and {Lee}, A. and {Lenz}, D. and {Littlefair}, S.~P. and {Ma}, Z. and {Macleod}, D.~M. and {Mastropietro}, M. and {McCully}, C. and {Montagnac}, S. and {Morris}, B.~M. and {Mueller}, M. and {Mumford}, S.~J. and {Muna}, D. and {Murphy}, N.~A. and {Nelson}, S. and {Nguyen}, G.~H. and {Ninan}, J.~P. and {N{\"o}the}, M. and {Ogaz}, S. and {Oh}, S. and {Parejko}, J.~K. and {Parley}, N. and {Pascual}, S. and {Patil}, R. and {Patil}, A.~A. and {Plunkett}, A.~L. and {Prochaska}, J.~X. and {Rastogi}, T. and {Reddy Janga}, V. and {Sabater}, J. and {Sakurikar}, P. and {Seifert}, M. and {Sherbert}, L.~E. and {Sherwood-Taylor}, H. and {Shih}, A.~Y. and {Sick}, J. and {Silbiger}, M.~T. and {Singanamalla}, S. and {Singer}, L.~P. and {Sladen}, P.~H. and {Sooley}, K.~A. and {Sornarajah}, S. and {Streicher}, O. and {Teuben}, P. and {Thomas}, S.~W. and {Tremblay}, G.~R. and {Turner}, J.~E.~H. and {Terr{\'o}n}, V. and {van Kerkwijk}, M.~H. and {de la Vega}, A. and {Watkins}, L.~L. and {Weaver}, B.~A. and {Whitmore}, J.~B. and {Woillez}, J. and {Zabalza}, V. and {Astropy Contributors}},
        title = "{The Astropy Project: Building an Open-science Project and Status of the v2.0 Core Package}",
      journal = {\aj},
         year = 2018,
        month = sep,
       volume = {156},
       number = {3},
          eid = {123},
        pages = {123},
          doi = {10.3847/1538-3881/aabc4f},
archivePrefix = {arXiv},
       eprint = {1801.02634},
 primaryClass = {astro-ph.IM},
       adsurl = {https://ui.adsabs.harvard.edu/abs/2018AJ....156..123A}
}

@ARTICLE{2022ApJ...935..167A,
       author = {{Astropy Collaboration} and {Price-Whelan}, Adrian M. and {Lim}, Pey Lian and {Earl}, Nicholas and {Starkman}, Nathaniel and {Bradley}, Larry and {Shupe}, David L. and {Patil}, Aarya A. and {Corrales}, Lia and {Brasseur}, C.~E. and {N{\"o}the}, Maximilian and {Donath}, Axel and {Tollerud}, Erik and {Morris}, Brett M. and {Ginsburg}, Adam and {Vaher}, Eero and {Weaver}, Benjamin A. and {Tocknell}, James and {Jamieson}, William and {van Kerkwijk}, Marten H. and {Robitaille}, Thomas P. and {Merry}, Bruce and {Bachetti}, Matteo and {G{\"u}nther}, H. Moritz and {Aldcroft}, Thomas L. and {Alvarado-Montes}, Jaime A. and {Archibald}, Anne M. and {B{\'o}di}, Attila and {Bapat}, Shreyas and {Barentsen}, Geert and {Baz{\'a}n}, Juanjo and {Biswas}, Manish and {Boquien}, M{\'e}d{\'e}ric and {Burke}, D.~J. and {Cara}, Daria and {Cara}, Mihai and {Conroy}, Kyle E. and {Conseil}, Simon and {Craig}, Matthew W. and {Cross}, Robert M. and {Cruz}, Kelle L. and {D'Eugenio}, Francesco and {Dencheva}, Nadia and {Devillepoix}, Hadrien A.~R. and {Dietrich}, J{\"o}rg P. and {Eigenbrot}, Arthur Davis and {Erben}, Thomas and {Ferreira}, Leonardo and {Foreman-Mackey}, Daniel and {Fox}, Ryan and {Freij}, Nabil and {Garg}, Suyog and {Geda}, Robel and {Glattly}, Lauren and {Gondhalekar}, Yash and {Gordon}, Karl D. and {Grant}, David and {Greenfield}, Perry and {Groener}, Austen M. and {Guest}, Steve and {Gurovich}, Sebastian and {Handberg}, Rasmus and {Hart}, Akeem and {Hatfield-Dodds}, Zac and {Homeier}, Derek and {Hosseinzadeh}, Griffin and {Jenness}, Tim and {Jones}, Craig K. and {Joseph}, Prajwel and {Kalmbach}, J. Bryce and {Karamehmetoglu}, Emir and {Ka{\l}uszy{\'n}ski}, Miko{\l}aj and {Kelley}, Michael S.~P. and {Kern}, Nicholas and {Kerzendorf}, Wolfgang E. and {Koch}, Eric W. and {Kulumani}, Shankar and {Lee}, Antony and {Ly}, Chun and {Ma}, Zhiyuan and {MacBride}, Conor and {Maljaars}, Jakob M. and {Muna}, Demitri and {Murphy}, N.~A. and {Norman}, Henrik and {O'Steen}, Richard and {Oman}, Kyle A. and {Pacifici}, Camilla and {Pascual}, Sergio and {Pascual-Granado}, J. and {Patil}, Rohit R. and {Perren}, Gabriel I. and {Pickering}, Timothy E. and {Rastogi}, Tanuj and {Roulston}, Benjamin R. and {Ryan}, Daniel F. and {Rykoff}, Eli S. and {Sabater}, Jose and {Sakurikar}, Parikshit and {Salgado}, Jes{\'u}s and {Sanghi}, Aniket and {Saunders}, Nicholas and {Savchenko}, Volodymyr and {Schwardt}, Ludwig and {Seifert-Eckert}, Michael and {Shih}, Albert Y. and {Jain}, Anany Shrey and {Shukla}, Gyanendra and {Sick}, Jonathan and {Simpson}, Chris and {Singanamalla}, Sudheesh and {Singer}, Leo P. and {Singhal}, Jaladh and {Sinha}, Manodeep and {Sip{\H{o}}cz}, Brigitta M. and {Spitler}, Lee R. and {Stansby}, David and {Streicher}, Ole and {{\v{S}}umak}, Jani and {Swinbank}, John D. and {Taranu}, Dan S. and {Tewary}, Nikita and {Tremblay}, Grant R. and {de Val-Borro}, Miguel and {Van Kooten}, Samuel J. and {Vasovi{\'c}}, Zlatan and {Verma}, Shresth and {de Miranda Cardoso}, Jos{\'e} Vin{\'\i}cius and {Williams}, Peter K.~G. and {Wilson}, Tom J. and {Winkel}, Benjamin and {Wood-Vasey}, W.~M. and {Xue}, Rui and {Yoachim}, Peter and {Zhang}, Chen and {Zonca}, Andrea and {Astropy Project Contributors}},
        title = "{The Astropy Project: Sustaining and Growing a Community-oriented Open-source Project and the Latest Major Release (v5.0) of the Core Package}",
      journal = {\apj},
         year = 2022,
        month = aug,
       volume = {935},
       number = {2},
          eid = {167},
        pages = {167},
          doi = {10.3847/1538-4357/ac7c74},
archivePrefix = {arXiv},
       eprint = {2206.14220},
 primaryClass = {astro-ph.IM},
       adsurl = {https://ui.adsabs.harvard.edu/abs/2022ApJ...935..167A}
}
\bibliographystyle{aasjournal}

\end{document}